\documentclass[
 amsmath,amssymb,
 aps, prx,
]{revtex4-2}

\usepackage{graphicx}
\usepackage{dcolumn}
\usepackage{bm}
\usepackage{comment}
\usepackage{multirow}
\usepackage{booktabs}
\usepackage{hyperref}

\begin{document}


\title{\textbf{Statistical Parameter Calibration via the Generalized Fluctuation Dissipation Theorem and Generative Modeling} 
}%

\author{Ludovico T. Giorgini}
\email{ludogio@mit.edu}
\homepage{https://ludogiorgi.github.io/}
\affiliation{Department of Mathematics, Massachusetts Institute of Technology, Cambridge, MA 02139, USA}

\author{Tobias Bischoff}
\affiliation{Aeolus Labs, San Francisco, CA 94107, USA}

\author{Andre N. Souza}
\homepage{https://sandreza.github.io/}
\affiliation{Department of Earth, Atmospheric and Planetary Sciences, Massachusetts Institute of Technology, Cambridge, MA 02139, USA}

\begin{abstract}
\noindent We introduce a response-theoretic framework that recasts parameter calibration of ergodic stochastic differential equations as a fluctuation-dissipation problem. Our central result is that the full Jacobian of any stationary observable with respect to drift and diffusion parameters admits an exact linear-response representation as a time-correlation integral evaluated along unperturbed dynamics alone, without perturbed simulations, adjoint derivations, or tangent-linear models. The key idea is to interpret infinitesimal parameter variations as causal perturbations of the dynamics, thereby bringing the Generalized Fluctuation-Dissipation Theorem to inverse problems. The resulting kernels couple observables to the score of the invariant density, for which modern score-estimation methods provide practical non-Gaussian estimators. We validate the framework across a hierarchy of models, from analytically tractable processes to stochastic parameterization in the chaotic Lorenz-96 system, and show that a single baseline trajectory can recover parameter sensitivities and calibration updates with accuracy comparable to finite-difference approaches. More broadly, the framework opens a new route to model calibration, statistical inverse problems, and uncertainty quantification when the quantities of interest are long-time statistics of complex dynamical systems.

\end{abstract}

\keywords{parameter calibration, fluctuation--dissipation theorem, score function, stochastic differential equations, inverse problems, generative modeling}
\maketitle

\section{\label{sec:intro} Introduction}

Calibrating the parameters of complex dynamical models so that model-implied statistics agree with empirical observations is a central task across the physical, engineering, and life sciences \cite{Tarantola2005,KennedyOHagan2001, nanda2023calibration, dunbar2021calibration, chis2011bayesian}. Formally, this is an inverse problem: given data, infer parameter values that render the model statistically consistent with measured observables (means, variances, spectra, exceedance probabilities, \emph{etc.}). These problems are typically ill-posed---multiple parameter sets can explain the data and small perturbations in observations may induce large changes in inferred parameters---and thus require principled regularization and uncertainty quantification \cite{Stuart2010,DashtiStuart2017}. A complementary challenge is computational: each evaluation of the forward map often entails integrating high-dimensional, stiff, or chaotic dynamics, making naïve search or repeated finite-difference sensitivity analyses prohibitively expensive \cite{Bellman1961,Chakrabarty2021,Dunbar2022}.

A broad repertoire of methodologies has emerged to address this problem. Likelihood-based estimators and Bayesian formulations provide principled inference and uncertainty quantification but frequently demand many forward solves and careful regularization \cite{Stuart2010,DashtiStuart2017}. Model-discrepancy formulations account for structural imperfections in the simulator at the cost of increased inferential complexity \cite{KennedyOHagan2001}. Gradient-based strategies---adjoint/variational methods, Gauss--Newton, and Levenberg--Marquardt---can be highly efficient when sensitivities are available, but deriving and maintaining adjoints for large codes is onerous and linearized sensitivities may be unreliable in strongly nonlinear regimes \cite{Lettermann2024}.

Approaches that avoid explicit gradient computation face their own trade-offs. Derivative-free, covariance-informed updates such as Ensemble Kalman Inversion (EKI) parallelize naturally, yet they typically return point estimates with limited posterior characterization and can struggle in multimodal or strongly nonlinear settings \cite{Iglesias2013,Schillings2017}. Fully probabilistic samplers (MCMC) \cite{Hastings1970,Gelman2013,Roberts2001,Cui2009, uq_kpp} deliver gold-standard uncertainty quantification, but their cost scales poorly when each likelihood evaluation requires an expensive time integration; hybrid \emph{calibrate--emulate--sample} pipelines alleviate this by combining fast optimization, surrogate modeling, and sampling \cite{Cleary2021}. Variational Bayesian approaches further reduce costs by optimizing tractable posterior approximations in large-scale inverse problems \cite{Povala2022}. When likelihoods are intractable, likelihood-free methods (e.g., ABC and related simulation-based inference) broaden applicability at the price of large simulation budgets \cite{Daher2025}.

In this work, we introduce a conceptually different route to parameter calibration by recasting it as a response-theoretic problem. We show that the sensitivity of any stationary observable to model parameters is governed by the same fluctuation--dissipation structure that underlies nonequilibrium response theory. The key idea is that an infinitesimal change in a drift or diffusion parameter can be interpreted as a causal perturbation of the dynamics, thereby bringing the Generalized Fluctuation--Dissipation Theorem (GFDT) \cite{Prost2009,cooper2011climate,baldovin2020understanding,ghil2020physics,majda_climate_response,MajdaBook,SantosGutierrezLucarini2022, LucariniGutierrezMoroneyZagli2026, ZagliColbrookLucariniMezicMoroney2026} directly to bear on parameter inverse problems. This leads to an exact linear-order identity: each entry of the Jacobian $\partial \langle \mathcal{A} \rangle / \partial \boldsymbol{\theta}$ of a stationary observable $\mathcal{A}$ can be written as a temporal correlation integral evaluated along the unperturbed dynamics. Because the full Jacobian is obtained from a single baseline trajectory at fixed parameter values, the resulting calibration strategy avoids repeated perturbed simulations and scales much more favorably with parameter dimension than finite-difference, tangent-linear, or adjoint-based approaches.

A longstanding obstacle to using the GFDT as a practical computational tool is that its response formulas involve the score function, that is, the gradient of the logarithm of the invariant density. Except in special low-dimensional settings, this quantity is rarely available in closed form and has historically been difficult to estimate accurately from data in nonlinear, high-dimensional systems. For this reason, despite its central role in nonequilibrium statistical physics, the GFDT has seen only limited direct computational use beyond Gaussian or near-Gaussian regimes. Recent advances in score estimation and generative modeling have changed this picture by providing practical data-driven estimators of the invariant-density score in high-dimensional settings \cite{song2021score,hyvarinen2005estimation,vincent2011connection}. Recent works have further shown that such estimators can be incorporated into the GFDT framework to recover accurate response predictions \cite{Giorgini2024,giorgini2025predicting}. These developments provide the missing ingredient needed to make response-theoretic calculations feasible in nonlinear, non-Gaussian systems.

Building on this advance, the present paper establishes a new connection. Rather than using the GFDT to predict the response to prescribed external forcing, we target parameter derivatives of stationary observables, the quantities that drive gradient-based calibration, inverse problems, and uncertainty quantification. We derive exact identities expressing the full Jacobian of stationary statistics with respect to both drift and diffusion parameters as GFDT correlation integrals, and we show that these identities lead to a practical, nonintrusive calibration framework. The method requires only a single baseline simulation: the resulting trajectory is used to estimate both the score and the full statistical Jacobian, which can then be used directly in regularized Gauss--Newton or Levenberg--Marquardt updates. We validate the approach on a hierarchy of models, ranging from analytically tractable processes to stochastic parameterization in the spatiotemporally chaotic Lorenz--96 system, and show that it achieves accuracy comparable to finite-difference references at a fraction of their simulation cost.

The paper is organized as follows. Section~\ref{sec:Motivation} formulates the parameter calibration problem and introduces the GFDT framework for computing statistical Jacobians. Section~\ref{sec:j_results} demonstrates the approach with analytical and computational examples, including an Ornstein--Uhlenbeck process and a quartic potential system. Section~\ref{sec:o_results} presents a unified set of calibration results, from low-dimensional stochastic climate models to stochastic parameterization in the spatiotemporally chaotic Lorenz--96 system. Section~\ref{sec:conclusions} discusses conclusions and future directions.

\section{\label{sec:Motivation} Motivation and Problem Formulation}

Consider the It\^o stochastic differential equation (SDE) for a $d$-dimensional state $\bm{x}_t$ driven by an $m$-dimensional Wiener process $\bm{W}_t$,
\begin{equation}
  d\bm{x}_t
  = \bm{F}(\bm{x}_t;\,\bm{\alpha})\,dt
  + \bm{\Sigma}(\bm{x}_t;\,\bm{\beta})\,d\bm{W}_t,
  \label{eq:sde_motivation}
\end{equation}
where $\bm{F}:\mathbb{R}^d\to\mathbb{R}^d$ is the drift, $\bm{\Sigma}:\mathbb{R}^d\to\mathbb{R}^{d\times m}$ is the diffusion factor, and the parameters are partitioned into drift parameters $\bm{\alpha}$ and diffusion parameters $\bm{\beta}$. When we treat them jointly we write $\bm{\theta}=(\bm{\alpha},\bm{\beta})$.

For clarity of exposition we assume that~\eqref{eq:sde_motivation} is ergodic with a unique invariant (steady-state) density $\rho_{\bm{\theta}}(\bm{x})$. Extensions to non-stationary or cyclo-stationary settings are possible and are used later in the response-theoretic developments (see App.~\ref{sec:derivation_gfdt}). The ergodicity assumption guarantees that time averages along a single long trajectory converge to ensemble averages, which is the practical basis for the estimators developed below. For any observable $\mathcal{A}:\mathbb{R}^d\to\mathbb{R}$ that is integrable with respect to $\rho_{\bm{\theta}}$, we denote the ensemble expectation under the invariant law by
\begin{equation}
  \langle \mathcal{A} \rangle_{\bm{\theta}}
  \;\equiv\;
  \int_{\mathbb{R}^d} \mathcal{A}(\bm{x})\,\rho_{\bm{\theta}}(\bm{x})\,d\bm{x}.
\end{equation}

Our calibration objective is to determine parameters $\bm{\theta}$ such that the model-implied statistics $\langle \mathcal{A}_i \rangle_{\bm{\theta}}$ match empirical targets $A_i$ inferred from data, for a chosen collection of observables $\{\mathcal{A}_i\}_{i=1}^{q}$. In other words, we aim for \emph{statistical} fidelity of the model rather than pathwise agreement of individual realizations. This viewpoint is natural in many applications---including climate and geophysical modeling, neuroscience, quantitative finance, and systems biology---where design, risk, or scientific inference depends on moments, spectra, or tail probabilities rather than on single trajectories. The following subsections formalize this objective as a regularized nonlinear least-squares problem, introduce the Generalized Fluctuation--Dissipation Theorem (GFDT) as a tool for computing the required parameter sensitivities, and describe the score estimation methods that make the approach practical.

\subsection{Nonlinear Least Squares}

Given a collection of observables $\{\mathcal{A}_i\}_{i=1}^{q}$, we assemble their stationary expectations into a single vector-valued forward map that sends parameters to statistics:
\begin{equation}
\bm{\mathcal{G}}(\bm{\theta})
=\big(\mathcal{G}_{\mathcal{A}_1}(\bm{\theta}),\ldots,\mathcal{G}_{\mathcal{A}_q}(\bm{\theta})\big)^\top
\in \mathbb{R}^q,
\label{eq:forward_map}
\end{equation}
where each component is the stationary expectation of the corresponding observable,
\begin{equation}
\mathcal{G}_{\mathcal{A}_i}(\bm{\theta})=\langle \mathcal{A}_i\rangle_{\bm{\theta}}.
\label{eq:forward_map_component}
\end{equation}
The map $\bm{\mathcal{G}}$ encodes all the statistical information we wish to reproduce. In a typical application, its components might include means, variances, covariances, or tail exceedance probabilities of physically relevant quantities.

Let $\bm{A}=(A_1,\ldots,A_q)^\top$ denote the vector of empirically observed target statistics. We measure the discrepancy between the model-implied and target statistics through a weighted least-squares objective. Introducing a symmetric positive-definite weighting matrix $\bm{B}\in\mathbb{R}^{q\times q}$ (often chosen as the identity or as an inverse covariance of the observables) and a regularization term $\mathcal{R}(\bm{\theta})$ that enforces well-posedness, the parameter calibration task reads
\begin{equation}
\min_{\bm{\theta}\in\mathbb{R}^p}\;
\mathcal{L}(\bm{\theta})
=\frac{1}{2}\,(\bm{\mathcal{G}}(\bm{\theta})-\bm{A})^\top
\bm{B}\,(\bm{\mathcal{G}}(\bm{\theta})-\bm{A})
+\mathcal{R}(\bm{\theta}),
\label{eq:nlls_objective}
\end{equation}
where $p$ denotes the dimension of the parameter vector $\bm{\theta}$. The first term penalizes the mismatch between model and data, while the regularizer prevents overfitting and stabilizes the inversion when the problem is ill-conditioned.

\medskip
\noindent\textbf{Gradient of the loss.}
To solve~\eqref{eq:nlls_objective} with gradient-based methods, we need the derivative of $\mathcal{L}$ with respect to $\bm{\theta}$. Define the residual vector
$\bm{r}(\bm{\theta})=\bm{\mathcal{G}}(\bm{\theta})-\bm{A}$,
which measures the componentwise mismatch between model statistics and targets. Let
\begin{equation}
\bm{S}(\bm{\theta})=\frac{\partial \bm{\mathcal{G}}}{\partial \bm{\theta}}(\bm{\theta})
\in\mathbb{R}^{q\times p}
\label{eq:jacobian_definition}
\end{equation}
denote the sensitivity matrix, i.e., the Jacobian of the forward map with respect to the parameters. Its $(i,j)$-entry records how the $i$-th statistic responds to a small change in the $j$-th parameter. In terms of these quantities, the gradient of the loss takes the standard form
\begin{equation}
\nabla_{\bm{\theta}}\mathcal{L}(\bm{\theta})
=\bm{S}(\bm{\theta})^\top \bm{B}\,\bm{r}(\bm{\theta})
+\nabla_{\bm{\theta}}\mathcal{R}(\bm{\theta}).
\label{eq:nlls_gradient}
\end{equation}
The first term projects the weighted residual back into parameter space through the transpose of the sensitivity matrix; the second adds the contribution from the regularizer.

\medskip
\noindent\textbf{Local linearization.}
Rather than minimizing~\eqref{eq:nlls_objective} globally (which would require many evaluations of the nonlinear map $\bm{\mathcal{G}}$), we proceed iteratively by linearizing the forward map around a current estimate $\bm{\theta}^\ast$:
\begin{equation}
\bm{\mathcal{G}}(\bm{\theta})
\approx \bm{\mathcal{G}}(\bm{\theta}^\ast)
+ \bm{S}(\bm{\theta}^\ast)\,(\bm{\theta}-\bm{\theta}^\ast).
\label{eq:forward_taylor}
\end{equation}
This first-order Taylor expansion reduces the nonlinear problem to a sequence of linear ones. Choosing a quadratic regularizer centered on the current iterate,
\begin{equation}
\mathcal{R}(\bm{\theta})
=\frac{1}{2}(\bm{\theta}-\bm{\theta}^\ast)^\top
\bm{\Gamma}\,(\bm{\theta}-\bm{\theta}^\ast),
\label{eq:quadratic_reg}
\end{equation}
with $\bm{\Gamma}\succeq 0$ a symmetric positive-semidefinite regularization matrix, and substituting the linearization~\eqref{eq:forward_taylor} into the loss~\eqref{eq:nlls_objective}, we obtain the Gauss--Newton (or damped Newton) system for the parameter increment $\bm{\vartheta}=\bm{\theta}-\bm{\theta}^\ast$:
\begin{equation}
\big(\bm{S}^\top \bm{B}\,\bm{S}+\bm{\Gamma}\big)\,\bm{\vartheta}
=\bm{S}^\top \bm{B}\,(\bm{A}-\bm{\mathcal{G}}(\bm{\theta}^\ast)).
\label{eq:gn_update}
\end{equation}
The left-hand side is a regularized normal matrix that is always invertible when $\bm{\Gamma}\succ 0$; the right-hand side is the sensitivity-weighted data misfit evaluated at the current parameter estimate. Equation~\eqref{eq:gn_update} can be solved directly for low-dimensional parameter vectors, while in large-scale problems iterative solvers such as conjugate gradients are typically employed. The regularization matrix $\bm{\Gamma}$ provides numerical stability and encodes prior information on the parameters.

\medskip
\noindent\textbf{Computational challenge.}
The entire iterative scheme hinges on the ability to evaluate the sensitivity matrix $\bm{S}$ in \eqref{eq:jacobian_definition}. Standard approaches compute $\bm{S}$ by finite differences (requiring $2p$ perturbed simulations for central differences), tangent linear models, or adjoint methods, all of which become expensive or impractical for chaotic, high-dimensional systems. In the following subsections we show how the Generalized Fluctuation--Dissipation Theorem (GFDT) yields a mathematically exact expression for $\bm{S}$ using only time-correlation functions from a single unperturbed trajectory. This eliminates the need for repeated simulations or adjoint derivations and provides a principled route to statistical parameter calibration.

\subsection{The Generalized Fluctuation--Dissipation Theorem (GFDT)}
\label{subsec:gfdt}

The Generalized Fluctuation--Dissipation Theorem (GFDT) provides the first-order (linear) change in the expectation of an observable $\mathcal{A}$ induced by small, causal perturbations to the drift and/or diffusion of the SDE. A full derivation valid for non-stationary baselines is given in App.~\ref{sec:derivation_gfdt}; here we specialize to perturbations about a stationary reference process.

To set up the perturbation framework, we augment the baseline SDE~\eqref{eq:sde_motivation} with small time-dependent corrections to both the drift and diffusion:
\begin{equation}
d\bm{x}_t
=\Big[\bm{F}(\bm{x}_t;\bm{\alpha})+\varepsilon\,\bm{\Psi}(\bm{x}_t,t)\Big]\,dt
+\Big[\bm{\Sigma}(\bm{x}_t;\bm{\beta})+\varepsilon\,\bm{\Lambda}(\bm{x}_t,t)\Big]\,d\bm{W}_t,
\label{eq:gfdt_perturbed_sde}
\end{equation}
with It\^o interpretation. Here $\varepsilon$ is a small bookkeeping parameter that tracks the perturbation order, $\bm{\Psi}:\mathbb{R}^d\times\mathbb{R}\to\mathbb{R}^d$ is the drift perturbation, and $\bm{\Lambda}:\mathbb{R}^d\times\mathbb{R}\to\mathbb{R}^{d\times m}$ is the diffusion perturbation. Both perturbations are assumed causal, meaning they vanish for negative times: $\bm{\Psi}(\cdot,t)=\bm{0}$ and $\bm{\Lambda}(\cdot,t)=\bm{0}$ for $t<0$. This ensures that the system starts from the unperturbed stationary state.

The central object in the GFDT is the \emph{score function} of the unperturbed invariant density,
\begin{equation}
\bm{s}(\bm{x}) \;\equiv\; \nabla_{\bm{x}}\!\log \rho(\bm{x}),
\label{eq:score_def}
\end{equation}
which is the gradient of the log-density with respect to the state variables. The score characterizes the local geometry of the invariant measure: it points in the direction of steepest increase of the probability density, and its magnitude reflects how sharply the density varies. The GFDT expresses the linear response of any observable in terms of correlations between that observable and a conjugate quantity built from the perturbation and the score.

Concretely, the GFDT asserts that the first-order change in the expectation of $\mathcal{A}$ at time $t$ is
\begin{equation}
\delta\!\left\langle \mathcal{A} \right\rangle(t)
=\varepsilon\int_{0}^{t}\!\mathcal{K}(t,s)\,ds,
\label{eq:gfdt_response_integral}
\end{equation}
where the integration runs from the onset of the perturbation at $s=0$ to the observation time $t$. The integrand $\mathcal{K}(t,s)$ is the response kernel, defined as a two-time correlation under the \emph{unperturbed} dynamics:
\begin{equation}
\mathcal{K}(t,s)
=-\,\Big\langle \mathcal{A}(\bm{x}_t)\,\mathcal{B}(\bm{x}_s,s)\Big\rangle.
\label{eq:gfdt_kernel_def}
\end{equation}
This correlation links the observable $\mathcal{A}$ evaluated at the later time $t$ with a conjugate observable $\mathcal{B}$ evaluated at the earlier time $s$, both computed along trajectories of the \emph{unperturbed} system. The minus sign is a convention that makes the response kernel positive for dissipative perturbations.

The conjugate observable $\mathcal{B}$ encodes how the perturbation couples to the invariant measure through the score:
\begin{equation}
\mathcal{B}(\bm{x},t)
=\nabla_{\bm{x}}\!\cdot \,\widetilde{\bm{\Psi}}(\bm{x},t)
+\widetilde{\bm{\Psi}}(\bm{x},t)\cdot \bm{s}(\bm{x}).
\label{eq:gfdt_B_def}
\end{equation}
The first term is the divergence of an effective perturbation $\widetilde{\bm{\Psi}}$ (with $\nabla_{\bm{x}}\!\cdot \bm{v} = \sum_i \partial_{x_i} v_i$ for a vector field $\bm{v}$), and the second is the inner product of that perturbation with the score. Together, these two contributions capture how the perturbation locally compresses or expands probability (through the divergence) and how it pushes probability along the gradient of the log-density (through the score coupling).

The effective perturbation $\widetilde{\bm{\Psi}}$ itself combines the direct drift perturbation with an additional drift-like contribution arising from changes in the diffusion:
\begin{equation}
\widetilde{\bm{\Psi}}(\bm{x},t)
=\bm{\Psi}(\bm{x},t)
-\frac{1}{2}\,\nabla_{\bm{x}}\!\cdot\!\Big(\bm{\Sigma}\bm{\Lambda}^\top+\bm{\Lambda}\bm{\Sigma}^\top\Big)
-\frac{1}{2}\,\Big(\bm{\Sigma}\bm{\Lambda}^\top+\bm{\Lambda}\bm{\Sigma}^\top\Big)\bm{s}(\bm{x}).
\label{eq:psi_tilde_def}
\end{equation}
Here the divergence of the $d\times d$ matrix $\bm{M}=\bm{\Sigma}\bm{\Lambda}^\top+\bm{\Lambda}\bm{\Sigma}^\top$ is taken row-wise, producing a $d$-dimensional vector whose $i$-th component is $\sum_j \partial_{x_j} M_{ij}$. The two correction terms in~\eqref{eq:psi_tilde_def} arise because a change in diffusion alters the Fokker--Planck probability flux even without a direct change in the drift; the GFDT absorbs this effect into an equivalent drift perturbation $\widetilde{\bm{\Psi}}$.

Two important special cases deserve emphasis. For purely drift perturbations ($\bm{\Lambda}\equiv \bm{0}$), \eqref{eq:psi_tilde_def} reduces to $\widetilde{\bm{\Psi}}=\bm{\Psi}$ and the conjugate observable~\eqref{eq:gfdt_B_def} depends only on the score $\bm{s}$. This is the simplest and most commonly encountered case. When diffusion perturbations are also present, evaluating~\eqref{eq:psi_tilde_def} requires the row-wise divergence of $\bm{\Sigma}\bm{\Lambda}^\top+\bm{\Lambda}\bm{\Sigma}^\top$ and the score. Furthermore, computing the divergence of $\widetilde{\bm{\Psi}}$ inside~\eqref{eq:gfdt_B_def} will additionally require the score Jacobian $\nabla_{\bm{x}}\bm{s}(\bm{x})\in\mathbb{R}^{d\times d}$ whenever $\bm{\Sigma}$ or $\bm{\Lambda}$ depends on the state $\bm{x}$.

In summary, the GFDT expresses the first-order change of any admissible observable as an integral of a two-time correlation computed from the unperturbed dynamics alone. The practical requirements are: (i) the score $\bm{s}(\bm{x})$ of the invariant density, (ii) for diffusion perturbations, the score Jacobian $\nabla_{\bm{x}}\bm{s}(\bm{x})$, and (iii) empirical estimates of the correlation~\eqref{eq:gfdt_kernel_def} along a single baseline trajectory.

\subsection{GFDT and the Jacobian of the Forward Map}
\label{subsec:gfdt_jacobian}

We now show how a change in model parameters can be recast as a causal perturbation of the form treated by the GFDT, thereby connecting the response kernel to the sensitivity matrix $\bm{S}$ of the parameter-to-statistics map.

Consider the baseline SDE~\eqref{eq:sde_motivation} and suppose that at $t=0$ the parameters undergo a small step perturbation $\bm{\alpha}\mapsto\bm{\alpha}+\delta\bm{\alpha}$, $\bm{\beta}\mapsto\bm{\beta}+\delta\bm{\beta}$. Writing $\Theta(t)$ for the Heaviside step function ($\Theta(t)=1$ for $t\ge 0$, $\Theta(t)=0$ for $t<0$), the perturbed dynamics take the form
\begin{align}
d\bm{x}_t
&=\Big(\bm{F}(\bm{x}_t;\bm{\alpha})
+\big[\bm{F}(\bm{x}_t;\bm{\alpha}+\delta\bm{\alpha})-\bm{F}(\bm{x}_t;\bm{\alpha})\big]\Theta(t)\Big)\,dt
\label{eq:step_pert_sde_1}\\
&\quad+\Big(\bm{\Sigma}(\bm{x}_t;\bm{\beta})
+\big[\bm{\Sigma}(\bm{x}_t;\bm{\beta}+\delta\bm{\beta})-\bm{\Sigma}(\bm{x}_t;\bm{\beta})\big]\Theta(t)\Big)\,d\bm{W}_t.
\label{eq:step_pert_sde_2}
\end{align}
The structure is transparent: for $t<0$ the perturbed and unperturbed dynamics coincide, and for $t>0$ the system evolves under the new parameter values. The Heaviside function ensures that the perturbation is causal, matching the framework of Sec.~\ref{subsec:gfdt}.

To bring this into the GFDT form~\eqref{eq:gfdt_perturbed_sde}, we linearize the parameter-induced changes. Assuming $\delta\bm{\alpha}$ and $\delta\bm{\beta}$ are small, a first-order Taylor expansion gives
\begin{equation}
\varepsilon\,\bm{\Psi}(\bm{x},t)
=\sum_{i} \delta\alpha_i\,\bm{J}_i(\bm{x})\,\Theta(t),
\qquad
\varepsilon\,\bm{\Lambda}(\bm{x},t)
=\sum_{i} \delta\beta_i\,\bm{K}_i(\bm{x})\,\Theta(t),
\label{eq:psi_lambda_lin}
\end{equation}
where the parameter Jacobians of the drift and diffusion are defined as
\begin{equation}
\bm{J}_i(\bm{x})=\partial_{\alpha_i}\bm{F}(\bm{x};\bm{\alpha})\in\mathbb{R}^d,
\qquad
\bm{K}_i(\bm{x})=\partial_{\beta_i}\bm{\Sigma}(\bm{x};\bm{\beta})\in\mathbb{R}^{d\times m}.
\label{eq:J_K_defs}
\end{equation}
These quantities are derivatives with respect to \emph{parameters} (evaluated at the state $\bm{x}$), and should be distinguished from the spatial Jacobian of the score that appears in the diffusion-response formulas. Intuitively, $\bm{J}_i(\bm{x})$ tells us how the drift vector at state $\bm{x}$ changes when the $i$-th drift parameter is perturbed, and $\bm{K}_i(\bm{x})$ does the same for the diffusion matrix.

For the diffusion-parameter contributions, it is convenient to absorb the diffusion perturbation into an effective drift-like term, following the same construction as in~\eqref{eq:psi_tilde_def}. For each diffusion parameter $\beta_i$, define
\begin{equation}
\widetilde{\bm{K}}_i(\bm{x})
=-\tfrac{1}{2}\,\nabla_{\bm{x}}\!\cdot\!\big(\bm{\Sigma}\bm{K}_i^\top+\bm{K}_i\bm{\Sigma}^\top\big)
-\tfrac{1}{2}\,\big(\bm{\Sigma}\bm{K}_i^\top+\bm{K}_i\bm{\Sigma}^\top\big)\bm{s}(\bm{x}),
\label{eq:Ktilde_def}
\end{equation}
where the row-wise divergence convention from~\eqref{eq:psi_tilde_def} applies, and $\bm{s}(\bm{x})=\nabla_{\bm{x}}\log\rho(\bm{x})$ is the score of the unperturbed invariant density. This effective perturbation $\widetilde{\bm{K}}_i$ enters the GFDT through the same algebraic channel as a drift perturbation $\bm{J}_i$, providing a unified treatment of drift and diffusion sensitivities.

With these definitions in hand, we specialize the GFDT of Sec.~\ref{subsec:gfdt} to the step perturbations in \eqref{eq:psi_lambda_lin}. Because we are interested in the change of \emph{stationary} statistics rather than the transient response, we take the long-time limit $t\to\infty$ to allow the system to equilibrate under the perturbed parameters. This yields the entries of the sensitivity matrix $\bm{S}$:
\begin{align}
\frac{\partial \mathcal{G}_{\mathcal{A}}}{\partial \alpha_j}
&=\lim_{t\to\infty}
\left(-\int_{0}^{t}
\Big\langle
\mathcal{A}(\bm{x}_t)\,
\big[\nabla_{\bm{x}}\!\cdot \bm{J}_j(\bm{x}_s)
+\bm{J}_j(\bm{x}_s)\!\cdot\!\bm{s}(\bm{x}_s)\big]
\Big\rangle\,ds\right),
\label{eq:stat_jac_alpha_limit}\\[6pt]
\frac{\partial \mathcal{G}_{\mathcal{A}}}{\partial \beta_j}
&=\lim_{t\to\infty}
\left(-\int_{0}^{t}
\Big\langle
\mathcal{A}(\bm{x}_t)\,
\big[\nabla_{\bm{x}}\!\cdot \widetilde{\bm{K}}_j(\bm{x}_s)
+\widetilde{\bm{K}}_j(\bm{x}_s)\!\cdot\!\bm{s}(\bm{x}_s)\big]
\Big\rangle\,ds\right).
\label{eq:stat_jac_beta_limit}
\end{align}
The angle brackets $\langle\cdot\rangle$ denote expectation under the unperturbed stationary process, and the integrands are two-time correlations between the observable $\mathcal{A}$ at time $t$ and the GFDT conjugate at time $s$, computed along trajectories of the unperturbed system. The limit $t\to\infty$ exists provided the unperturbed dynamics are sufficiently mixing, so that the two-time correlations decay fast enough for the integral to converge.

Equations~\eqref{eq:stat_jac_alpha_limit} and~\eqref{eq:stat_jac_beta_limit} are the main results of this paper. They establish that the full parameter Jacobian of any stationary observable is determined by the unperturbed dynamics alone: each entry of the sensitivity matrix $\bm{S}=\partial \bm{\mathcal{G}}/\partial\bm{\theta}$ admits an explicit representation as a time-correlation integral, linking parameter sensitivities of statistics directly to the structural dependence of the drift and diffusion on the parameters in~\eqref{eq:sde_motivation}. The drift-parameter sensitivities in~\eqref{eq:stat_jac_alpha_limit} require only the parameter Jacobians $\bm{J}_j$ of the drift and the score $\bm{s}$. The diffusion-parameter sensitivities in~\eqref{eq:stat_jac_beta_limit} additionally involve the effective perturbation $\widetilde{\bm{K}}_j$ defined in~\eqref{eq:Ktilde_def}, which couples the diffusion parameter derivatives $\bm{K}_j$ to the score. If $\bm{\Sigma}$ or $\bm{K}_j$ depends on the state $\bm{x}$, evaluating the divergence of $\widetilde{\bm{K}}_j$ inside the integrand of~\eqref{eq:stat_jac_beta_limit} further requires the score Jacobian $\nabla_{\bm{x}}\bm{s}(\bm{x})\in\mathbb{R}^{d\times d}$.

\medskip
\noindent\textbf{Algorithmic summary.}
Given an observable $\mathcal{A}$ and baseline parameters $\bm{\theta}=(\bm{\alpha},\bm{\beta})$, the sensitivity matrix $\bm{S}$ is constructed as follows:
\begin{enumerate}
\item \textbf{Simulate} a single long trajectory (or ensemble) of the unperturbed system \eqref{eq:sde_motivation}. This trajectory serves as the statistical sample from the invariant measure. \label{step:simulate}
\item \textbf{Estimate the score} $\bm{s}(\bm{x})=\nabla_{\bm{x}}\log\rho(\bm{x})$ from the trajectory data. For diffusion-parameter sensitivities, also estimate the score Jacobian $\nabla_{\bm{x}}\bm{s}(\bm{x})$, for example via automatic differentiation of a differentiable score model. \label{step:score}
\item \textbf{Form the correlation integrals} in \eqref{eq:stat_jac_alpha_limit}--\eqref{eq:stat_jac_beta_limit} using the trajectory from step~\ref{step:simulate} and the score estimates from step~\ref{step:score}. In practice, the two-time correlations are evaluated as empirical averages over the trajectory, and the time integral is carried out numerically (e.g., via FFT-based correlation estimators). \label{step:correlations}
\end{enumerate}
The resulting sensitivity matrix $\bm{S}$ can be embedded within the Gauss--Newton or Levenberg--Marquardt loop of Sec.~\ref{subsec:gfdt_jacobian} to update $\bm{\theta}$ iteratively. Practical variants, including choices of integration windows, sampling cadences, and convergence criteria, are discussed in Sec.~\ref{sec:o_results}.

\subsection{Estimating the Score and Its Jacobian for GFDT Calibration}
\label{subsec:score_estimation}

The GFDT formulas in Eqs.~\eqref{eq:stat_jac_alpha_limit}--\eqref{eq:stat_jac_beta_limit} require the score
$\bm{s}(\bm{x})=\nabla_{\bm{x}}\log\rho(\bm{x})$ of the (unknown) stationary density and, for diffusion-parameter perturbations, its Jacobian $\nabla_{\bm{x}}\bm{s}(\bm{x})\in\mathbb{R}^{d\times d}$. Since the invariant density $\rho$ is generally not available in closed form, these quantities must be estimated from data. We adopt denoising score matching (DSM) as the primary estimation strategy, where a neural network is trained to approximate the score from noisy samples. For the low-dimensional systems considered in this work, we accelerate this training using the K-Means Gaussian Mixture Modeling (KGMM) algorithm \cite{giorgini2025kgmm}, which provides an efficient non-parametric estimator of the conditional expectation that DSM targets. Both DSM and its KGMM-accelerated variant yield a differentiable estimate of the score, from which we construct $\nabla_{\bm{x}}\bm{s}(\bm{x})$ using reverse-mode automatic differentiation. These quantities form the essential inputs for the GFDT-based calibration framework.

\medskip
\noindent\textbf{Denoising score matching (DSM).}
The key idea behind DSM is to avoid estimating the density $\rho$ directly and instead learn its score $\nabla_{\bm{x}}\log\rho$ by solving a regression problem. Given samples $\{\bm{x}^{(n)}\}_{n=1}^N\!\sim\!\rho$ from an unperturbed trajectory, DSM perturbs each sample by a small isotropic Gaussian noise and trains a neural network to recover the noise that was added. Specifically, for a fixed noise level $\sigma_{\mathrm{dsm}}>0$, draw corrupted samples
\begin{equation}
\bm{x}'=\bm{x}+\sigma_{\mathrm{dsm}}\,\bm{z},
\qquad
\bm{z}\sim\mathcal{N}(\bm{0},\bm{I}_d),
\label{eq:dsm_perturb}
\end{equation}
where $\bm{I}_d$ is the $d\times d$ identity matrix and $\bm{z}$ is a standard normal noise vector. A neural network $\widehat{\bm{g}}(\bm{x}')$ is then trained to predict $\bm{z}$ from $\bm{x}'$ by minimizing the mean-squared denoising objective
\begin{equation}
\mathcal{L}_{\mathrm{DSM}}
=\mathbb{E}\Big[\big\|\widehat{\bm{g}}(\bm{x}')-\bm{z}\big\|_2^2\Big],
\label{eq:dsm_loss_fixed_sigma}
\end{equation}
where the expectation is over the empirical data distribution and the corruption in~\eqref{eq:dsm_perturb}. The network that minimizes this loss is the Bayes-optimal denoiser,
\begin{equation}
\widehat{\bm{g}}^{\star}(\bm{x}')
=\mathbb{E}[\bm{z}\mid \bm{x}'],
\label{eq:dsm_bayes_optimal}
\end{equation}
which predicts the conditional mean of the noise given the corrupted sample. The connection to the score arises from a standard identity: the score of the noise-smoothed density $\rho_{\sigma}\equiv\rho\ast\mathcal{N}(0,\sigma_{\mathrm{dsm}}^2\bm{I}_d)$ satisfies
\begin{equation}
\nabla_{\bm{x}'}\log \rho_{\sigma}(\bm{x}')
=-\frac{1}{\sigma_{\mathrm{dsm}}}\,\mathbb{E}[\bm{z}\mid \bm{x}'].
\label{eq:dsm_score_identity}
\end{equation}
For small $\sigma_{\mathrm{dsm}}$, the smoothed density $\rho_{\sigma}$ is a close approximation to $\rho$, so the trained network provides the score estimator
\begin{equation}
\widehat{\bm{s}}(\bm{x})
\;\approx\;
-\frac{1}{\sigma_{\mathrm{dsm}}}\,\widehat{\bm{g}}(\bm{x}).
\label{eq:dsm_score_estimator}
\end{equation}
Because $\widehat{\bm{g}}$ is a differentiable neural network, reverse-mode automatic differentiation directly yields the score Jacobian $\nabla_{\bm{x}}\widehat{\bm{s}}(\bm{x})$ needed in the diffusion-parameter GFDT integrands.

\medskip
\noindent\textbf{Low-dimensional regime: KGMM estimator.}
When the effective state dimension is small, the conditional expectation in~\eqref{eq:dsm_bayes_optimal} can be computed non-parametrically by clustering, avoiding the cost of explicit neural network training. The idea is to partition the corrupted-sample space into regions and estimate $\mathbb{E}[\bm{z}\mid\bm{x}']$ within each region by a simple average.

Let $\{\bm{\mu}_i\}_{i=1}^N$ denote $N$ data samples drawn from the stationary distribution (here $\bm{\mu}_i$ labels the $i$-th sample, following the notation of \cite{giorgini2025kgmm}). Generate perturbed proxies with a kernel width $\sigma_G>0$:
\begin{equation}
\bm{x}_i=\bm{\mu}_i+\sigma_G\,\bm{z}_i,
\qquad
\bm{z}_i\sim\mathcal{N}(\bm{0},\bm{I}_d).
\label{eq:kgmm_perturb}
\end{equation}
Partition the perturbed samples $\{\bm{x}_i\}$ into $N_C$ clusters $\{\Omega_j\}_{j=1}^{N_C}$ (e.g., via a modified bisecting $k$-means, see \cite{souza2024modified}), and let $\bm{C}_j$ be the centroid of cluster $\Omega_j$. Within each cluster, the noise vectors $\bm{z}_i$ are approximately drawn from the same conditional distribution, so their average provides an estimate of the conditional mean in~\eqref{eq:dsm_bayes_optimal}:
\begin{equation}
\widehat{\mathbb{E}}[\bm{z}\mid \bm{x}\in\Omega_j]
=\frac{1}{|\Omega_j|}\sum_{\bm{x}_i\in\Omega_j}\bm{z}_i.
\label{eq:kgmm_cluster_conditional}
\end{equation}
Applying the same relationship between the conditional mean and the score as in~\eqref{eq:dsm_score_identity}, we obtain the score estimate at each centroid:
\begin{equation}
\widehat{\bm{s}}(\bm{C}_j)
\;=\;
-\frac{1}{\sigma_G}\,\widehat{\mathbb{E}}[\bm{z}\mid \bm{x}\in\Omega_j].
\label{eq:kgmm_score_on_centroids}
\end{equation}
The centroid-score pairs $\{(\bm{C}_j,\widehat{\bm{s}}(\bm{C}_j))\}$ are then interpolated with a smooth regressor (e.g., a multilayer perceptron) to obtain a differentiable score field $\widehat{\bm{s}}(\bm{x})$ defined throughout state space. Automatic differentiation of this interpolant yields the score Jacobian $\nabla_{\bm{x}}\widehat{\bm{s}}(\bm{x})$. The KGMM approach leverages clustering to evaluate $\mathbb{E}[\bm{z}\mid \bm{x}]$ without explicit mixture likelihoods, providing robust and computationally efficient score estimates in low dimensions. Implementation details for the calibration experiments are given in App.~\ref{app:calib_details}.

\section{\label{sec:j_results} Jacobian Results}

Before turning to full calibration loops, we first verify that the GFDT correlation formulas~\eqref{eq:stat_jac_alpha_limit}--\eqref{eq:stat_jac_beta_limit} recover correct parameter sensitivities on problems where independent reference values are available. We consider two testbeds of increasing complexity. First, for the Ornstein--Uhlenbeck (OU) process we compute the sensitivity of the stationary variance with respect to the damping parameter both by direct differentiation of the closed-form variance and via the GFDT, obtaining exact agreement. This provides a fully analytic consistency check. Second, for a nonlinear double-well (quartic) potential we evaluate Jacobians of several observables---including a discontinuous tail indicator---using (i) GFDT with the exact score, (ii) high-accuracy finite differences as a reference, and (iii) a quasi-Gaussian closure as a baseline approximation. The second example highlights the regimes where an accurate, non-Gaussian score is essential.

\subsection{\label{sec:j_OU} Analytic Warmup: Ornstein--Uhlenbeck Process}

We begin with the simplest nontrivial setting: the scalar Ornstein--Uhlenbeck process. In the notation of the general framework~\eqref{eq:sde_motivation}, the state is one-dimensional ($d=1$, $m=1$), the drift is $F(x;\alpha)=-\alpha x$ with a single drift parameter $\alpha>0$ (the damping rate), and the diffusion is $\Sigma(x;\beta)=\beta$ with a single diffusion parameter $\beta>0$ (the noise amplitude). The SDE reads
\begin{equation}
dx_t \;=\; -\alpha\,x_t\,dt \;+\; \beta\,dW_t .
\end{equation}
Because the drift is linear and the noise is additive, the stationary distribution is Gaussian and available in closed form. The invariant density is
\begin{equation}
\rho(x) = \sqrt{\frac{\alpha}{\pi\beta^2}}\,\exp\!\Big(-\frac{\alpha x^2}{\beta^2}\Big),
\end{equation}
so the stationary distribution is $\mathcal{N}(0,\,\beta^2/(2\alpha))$, with variance $\langle x^2\rangle = \beta^2/(2\alpha)$. The score function (gradient of the log-density) is
\begin{equation}
s(x) \;=\; \nabla\log\rho(x) \;=\; -\frac{2\alpha}{\beta^2}\,x.
\end{equation}
Since the process is Gaussian and stationary, its two-time covariance is an exponentially decaying function of the time lag:
\begin{equation}
\langle x_t x_s \rangle \;=\; \frac{\beta^2}{2\alpha}\,e^{-\alpha|t-s|}\!,
\qquad t,s\in\mathbb{R}.
\end{equation}

Our goal is to compute the sensitivity of the stationary variance $\langle x^2 \rangle = \beta^2/(2\alpha)$ with respect to the damping parameter $\alpha$ in two independent ways: (i) by direct differentiation of the analytic variance expression, and (ii) via the GFDT using steady-state time correlations of the unperturbed process. Agreement between the two would confirm that the GFDT machinery correctly recovers the parameter sensitivity.

\medskip
\noindent\textbf{Method 1 (direct differentiation).}
Since the variance is known in closed form as $\langle x^2 \rangle = \beta^2/(2\alpha)$, differentiation with respect to $\alpha$ gives
\begin{equation}
\frac{\partial}{\partial \alpha}\, \langle x^2 \rangle 
\;=\; -\,\frac{\beta^2}{2\alpha^2}.
\label{eq:ou-direct}
\end{equation}
This is the reference value against which the GFDT calculation will be compared.

\medskip
\noindent\textbf{Method 2 (GFDT).}
We now derive the same quantity using the drift-parameter GFDT formula~\eqref{eq:stat_jac_alpha_limit}. The observable is $\mathcal{A}(x)=x^2$, and the parameter Jacobian of the drift (as defined in Eq.~\eqref{eq:J_K_defs}) is the scalar function
\begin{equation}
J(x)=\partial_\alpha F(x;\alpha)=\partial_\alpha(-\alpha x)=-x.
\end{equation}
According to~\eqref{eq:gfdt_B_def}, the GFDT conjugate observable is built from the divergence of $J$ and its inner product with the score. In one dimension, these reduce to ordinary derivatives and products:
\begin{align}
\nabla\!\cdot J + J\cdot s(x)
&= \underbrace{\partial_x(-x)}_{=-1}
\;+\; \underbrace{(-x)\left(-\frac{2\alpha}{\beta^2}x\right)}_{=\frac{2\alpha}{\beta^2}x^2}
\;=\; -1 + \frac{2\alpha}{\beta^2}\,x^2.
\end{align}
The first term ($-1$) comes from the spatial divergence of the drift perturbation, and the second term couples the perturbation to the score. Incorporating the Heaviside step function $\Theta(s)$ to enforce causality (the perturbation is applied at $s=0$), the GFDT response kernel~\eqref{eq:gfdt_kernel_def} becomes
\begin{equation}
\mathcal{K}(t,s) \;=\; -\,\big\langle \mathcal{A}(x_t)\,\mathcal{B}(x_s, s)\big\rangle
\;=\; \left\langle x_t^2 \right\rangle\Theta(s)
- \frac{2\alpha}{\beta^2}\left\langle x_t^2 x_s^2 \right\rangle\Theta(s).
\end{equation}
For $s < 0$, the kernel vanishes because the perturbation has not yet been applied. For $s > 0$, we need the fourth-order moment $\langle x_t^2 x_s^2\rangle$. Since $x_t$ and $x_s$ are jointly Gaussian (being values of the stationary OU process at two different times), we can apply Isserlis' theorem (also known as Wick's theorem), which expresses higher-order moments of jointly Gaussian variables in terms of products of second-order moments:
\begin{equation}
\langle x_t^2 x_s^2\rangle
= \langle x_t^2\rangle\langle x_s^2\rangle + 2\langle x_t x_s\rangle^2.
\end{equation}
Substituting the known stationary moments $\langle x_t^2\rangle=\langle x_s^2\rangle=\beta^2/(2\alpha)$ and $\langle x_t x_s\rangle=\tfrac{\beta^2}{2\alpha} e^{-\alpha|t-s|}$ yields
\begin{align}
\left\langle x_t^2 x_s^2 \right\rangle
&= \left(\frac{\beta^2}{2\alpha}\right)^2 \!\left(1 + 2\,e^{-2\alpha|t-s|}\right).
\end{align}
Inserting this into the response kernel expression and simplifying, we obtain
\begin{align}
\mathcal{K}(t,s)
&= \frac{\beta^2}{2\alpha}
- \frac{2\alpha}{\beta^2}\left(\frac{\beta^2}{2\alpha}\right)^2\!\left(1 + 2\,e^{-2\alpha|t-s|}\right)
= -\,\frac{\beta^2}{2\alpha}\,e^{-2\alpha|t-s|},
\end{align}
for $s > 0$, and $\mathcal{K}(t,s) = 0$ otherwise. The constant terms cancel exactly, leaving only the exponentially decaying correlation. This cancellation is a consequence of the fluctuation--dissipation balance: the score coupling precisely subtracts the disconnected (mean-field) part of the four-point correlator, isolating the connected fluctuation structure that governs the linear response.

To extract the stationary parameter sensitivity, we integrate the kernel over the entire history of the step perturbation. This corresponds to applying a small step change $\alpha\mapsto \alpha+\varepsilon$ at $s=0$ and waiting for the system to reach its new steady state:
\begin{align}
\delta\langle x^2\rangle(t)
&= \varepsilon \int_{0}^{t}\mathcal{K}(t,s)\,ds
= -\,\varepsilon\frac{\beta^2}{2\alpha} \int_{0}^{t} e^{-2\alpha(t-s)}\,ds
= -\,\varepsilon\frac{\beta^2}{2\alpha^2}\left(1-e^{-2\alpha t}\right).
\end{align}
The factor $(1-e^{-2\alpha t})$ captures the transient adjustment from the old steady state to the new one. At finite $t$, the system has only partially equilibrated; the full stationary sensitivity is obtained by taking $t\to\infty$, which removes the transient:
\begin{equation}
\lim_{t\to\infty}\frac{\delta\langle x^2\rangle(t)}{\varepsilon}
\;=\; -\,\frac{\beta^2}{2\alpha^2}.
\label{eq:ou-gfdt}
\end{equation}

Comparing~\eqref{eq:ou-direct} and~\eqref{eq:ou-gfdt}, we see that the two methods yield identical results:
\begin{align}
\underbrace{\frac{\partial \langle x^2 \rangle}{\partial \alpha}}_{\text{direct differentiation}} 
= - \frac{\beta^2}{2 \alpha^2} 
= \underbrace{\lim_{t \rightarrow \infty} -\int_0^t ds \left\langle (x_t)^2 \left(\partial_{x_s}(-x_s) - x_s \nabla \ln \rho(x_s) \right) \right\rangle }_{ \text{GFDT correlation integral}}.
\end{align}
The GFDT recovers the exact derivative of the stationary variance using only statistics of the \emph{unperturbed} process. In practice, this means that the sensitivity could be estimated from a single long simulation at the baseline parameter values, with no need to re-simulate under perturbed parameters. While the OU example is simple enough that direct differentiation is possible, it serves as a proof of concept: the same GFDT correlation structure applies to nonlinear, non-Gaussian systems where closed-form expressions are unavailable.

\subsection{\label{sec:j_DW} Quartic Potential}

We now move beyond the Gaussian setting to a nonlinear system with a genuinely non-Gaussian invariant measure. Consider a particle in a quartic potential with tunable drift parameters $\bm{\alpha}=(\alpha_1,\alpha_2,\alpha_3,\alpha_4)$ and diffusion parameters $\bm{\beta}=(\beta_1,\beta_2,\beta_3)$:
\begin{equation}
dx_t = \left(\alpha_1 + \alpha_2 x_t + \alpha_3 x_t^2 + \alpha_4 x_t^3 \right)\,dt + \sqrt{(\beta_1 + \beta_2 x_t)^2 + \beta_3^2}\,dW_t.
\end{equation}
The drift is a general cubic polynomial in $x$, and the diffusion amplitude depends on $x$ through the parameters $\beta_1$ and $\beta_2$, with $\beta_3$ providing a state-independent floor that prevents the noise from vanishing. This model is rich enough to exhibit bimodal (double-well) structure, asymmetric tails, and state-dependent noise, while remaining one-dimensional so that reference solutions can be obtained by direct quadrature of the invariant density.

For a one-dimensional It\^o SDE of the form $dx = f(x)\,dt + g(x)\,dW_t$, the stationary Fokker--Planck equation can be integrated once to yield the invariant density in closed form (up to normalization). The corresponding score function is
\begin{equation}
s(x) = \nabla \ln \rho(x) = \frac{2\left(\alpha_1 + \alpha_2 x + \alpha_3 x^2 + \alpha_4 x^3\right)}{(\beta_1 + \beta_2 x)^2 +  \beta_3^2} - \frac{2\left(\beta_2^2 x +  \beta_1 \beta_2\right)}{(\beta_1 + \beta_2 x)^2 +  \beta_3^2}.
\end{equation}
The first term arises from the drift and the second from the state-dependent diffusion (it would vanish if the noise were purely additive).

\medskip
\noindent\textbf{Baseline parameters and simplified dynamics.}
For the baseline simulation, we choose $(\alpha_1, \alpha_2, \alpha_3, \alpha_4 ) = (0, 1, 0, -1)$ and $(\beta_1, \beta_2, \beta_3) = (1, 0, 1)$. With these values, the diffusion is constant ($\sqrt{1+1}=\sqrt{2}$) and the drift becomes a symmetric cubic, yielding the dynamics
\begin{equation}
dx_t = \left(x_t - x_t^3 \right)\,dt + \sqrt{2}\,dW_t,
\end{equation}
which describes a particle in the symmetric double-well potential $V(x)=-\tfrac{1}{2}x^2+\tfrac{1}{4}x^4$, driven by additive white noise. The score simplifies to
\begin{equation}
s(x) = x - x^3.
\end{equation}
This baseline has a bimodal invariant density with peaks near $x=\pm 1$ and is symmetric about the origin, so all odd moments vanish. The non-Gaussian character of this distribution (heavy tails, bimodality) makes it a useful testbed for assessing the importance of accurate score estimation.

\medskip
\noindent\textbf{Observables.}
We compute the parameter Jacobian of three observables that probe different aspects of the invariant measure. The first two are standard polynomial moments:
\begin{equation}
\mathcal{A}_1(x) = x \quad (\text{mean}), \qquad \mathcal{A}_2(x) = x^2 \quad (\text{second moment}).
\end{equation}
The third is a discontinuous indicator function that captures tail behavior:
\begin{equation}
\mathcal{A}_3(x) = \mathbf{1}_{x>2}  \equiv \begin{cases}
1 & \text{if } x > 2 \\
0 & \text{if } x \leq 2
\end{cases}.
\end{equation}
The expected value $\langle \mathcal{A}_3 \rangle = \Pr(x > 2)$ is the probability that the particle occupies the far tail of the distribution. Its parameter Jacobian quantifies how this exceedance probability shifts as the model parameters change. We include this observable as a proxy for how extreme statistics respond to parameter perturbations, a question of practical importance in climate and risk applications. The GFDT formulas~\eqref{eq:stat_jac_alpha_limit}--\eqref{eq:stat_jac_beta_limit} apply to discontinuous observables, since the derivation (App.~\ref{sec:derivation_gfdt}) requires integrability of $\mathcal{A}$ with respect to the invariant measure but not smoothness.

\begin{figure}
  \begin{center}
  \includegraphics[width=1.0\textwidth]{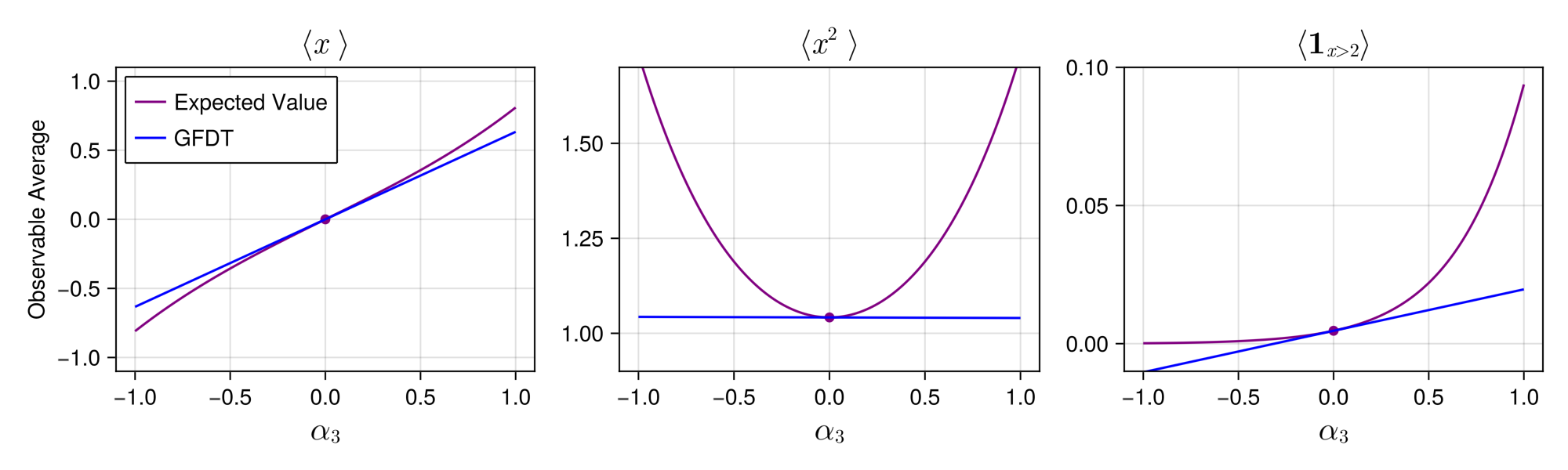}
  \caption[Parameter dependence in the quartic potential.]{\textbf{Parameter Dependence of Statistics in Quartic Potential System.} By varying the coefficient in front of the quadratic term (the $\alpha_3$ parameter) we get different values for the mean (left), second moment (middle), and the probability that $x > 2.0$ (right). The GFDT obtains the tangent line to this curve (blue) for the choice of parameters $\vec{\alpha} = (0,1,0,-1)$.}
  \end{center}
  \label{fig:gfdt_jacobian_2}
\end{figure}

\medskip
\noindent\textbf{Parameter dependence of statistics.}
Each observable's stationary expectation depends on the choice of parameters, and this dependence is generally nonlinear. To visualize this, Figure~\ref{fig:gfdt_jacobian_2} shows how the three statistics vary as the coefficient $\alpha_3$ (the quadratic term in the drift) is swept away from its baseline value of zero, while all other parameters are held fixed. The resulting curves (shown in purple) are computed by direct numerical quadrature of the one-dimensional invariant density at each parameter value, which is inexpensive for this scalar problem. 

The GFDT allows us to compute the tangent line (shown in blue) to each of these parameter-dependent curves at the baseline parameter values (the purple dot), using only a single simulation at the baseline. This tangent line is the local linear approximation $\langle \mathcal{A} \rangle(\alpha_3) \approx \langle \mathcal{A} \rangle(0) + (\partial \langle \mathcal{A} \rangle / \partial \alpha_3)\,\alpha_3$, where the slope is the GFDT-derived Jacobian entry. Figure~\ref{fig:gfdt_jacobian_2} illustrates three qualitatively different scenarios: a case where the linear approximation is accurate over a wide parameter range (the mean), a case where it captures the local slope well but the true dependence is strongly nonlinear (the second moment), and a case where the linear approximation remains useful locally despite the underlying nonlinearity (the tail probability). The extent to which the tangent line tracks the true curve depends on the degree of nonlinearity in the parameter-to-statistics map, which the GFDT does not attempt to capture beyond first order.

\medskip
\noindent\textbf{Quantitative comparison of Jacobians.}
Table~\ref{t:parameter_jacobian_quartic} reports a quantitative comparison of the full parameter Jacobian evaluated at the baseline $(\alpha_1,\alpha_2,\alpha_3,\alpha_4)=(0,1,0,-1)$. We compute the sensitivity of each observable with respect to each drift parameter using three methods. The first is high-accuracy finite differences: because the system is one-dimensional, the invariant density can be computed by numerical quadrature at closely spaced parameter values, and the resulting statistic curves can be differentiated numerically to high accuracy. We regard these finite-difference values as the reference. The second method is the GFDT applied to a long time series, using the exact analytic score $s(x)=x-x^3$. The third is a \emph{quasi-Gaussian} GFDT, which replaces the true invariant density with a Gaussian having the same mean and variance, and uses the corresponding Gaussian score in the GFDT formulas. This last calculation serves as a data-driven benchmark for the effect of score-model misspecification: it shows what happens when the non-Gaussian structure of the invariant measure is ignored.

\begin{table}[h!]
\centering
\caption[Comparison of Jacobian estimates.]{Comparison of Jacobian estimates. Each row corresponds to an observable, and each column to a perturbation of a model parameter. The Jacobians are computed using finite differences, the GFDT, and a Gaussian approximation to the GFDT. We regard the finite difference calculation as the best proxy to a ``ground truth'' in this particular case.}
\begin{tabular}{llcccc}
\toprule
\textbf{Method} & \textbf{Observable} & \textbf{\( \partial / \partial \alpha_1 \)} & \textbf{\( \partial / \partial\alpha_2 \)} & \textbf{\( \partial / \partial\alpha_3 \)} & \textbf{\( \partial / \partial \alpha_4 \)} \\
\midrule
\multirow{3}{*}{Finite differences} 
& \( \langle x \rangle \)      & 1.0418   & $-8.33\!\times\!10^{-14}$ & 0.6806  & $-2.78\!\times\!10^{-14}$ \\
& \( \langle x^2 \rangle \)    & 0.0      & 0.4782                    & 0.0     & $-0.7600$ \\
& \( \left\langle \mathbf{1}_{x>2} \right\rangle \)    & 0.00987  & 0.00804                   & 0.01484 & $-0.02134$ \\
\midrule
\multirow{3}{*}{Analytical} 
& \( \langle x \rangle \)      & 0.9244   & $-0.00398$                & 0.6403  & 0.01788 \\
& \( \langle x^2 \rangle \)    & $-0.01322$ & 0.5085                 & $-0.02305$ & $-0.9090$ \\
& \( \langle \mathbf{1}_{x>2} \rangle \)    & 0.00941  & 0.00858                   & 0.01548 & $-0.02369$ \\
\midrule
\multirow{3}{*}{GFDT: Quasi-Gaussian} 
& \( \langle x \rangle \)      & 1.1031   & 0.00153                   & $-0.2441$ & $-0.01072$ \\
& \( \langle x^2 \rangle \)    & $-0.00149$ & 0.3536                 & 0.00855   & 5.8803 \\
& \( \langle \mathbf{1}_{x>2} \rangle \)    & 0.00808  & 0.00354                   & 0.00088   & 0.02472 \\
\bottomrule
\end{tabular}
\label{t:parameter_jacobian_quartic}
\end{table}

The $\partial/\partial\alpha_3$ column of Table~\ref{t:parameter_jacobian_quartic} corresponds to the slopes of the blue tangent lines in Figure~\ref{fig:gfdt_jacobian_2}. The GFDT with the analytic score closely reproduces the finite-difference reference slopes across all observables and parameter directions. The residual discrepancies (e.g., small nonzero values where exact symmetry predicts zero) arise from finite trajectory length in the empirical correlation estimates.

The quasi-Gaussian approximation tells a different story. For some entries, particularly the $\alpha_1$ sensitivities of low-order moments, it performs reasonably: this is expected because the Gaussian score captures the leading-order (linear) response structure, and the $\alpha_1$ perturbation (a constant shift in the drift) does not strongly probe the tails. However, the quasi-Gaussian closure predicts the wrong sign for the mean response to $\alpha_3$ (the quadratic term in the drift) and the wrong order of magnitude for the tail observable $\langle \mathbf{1}_{x>2}\rangle$ with respect to $\alpha_4$ (the cubic term in the drift), where it gives $+0.025$ instead of the correct $-0.021$. These failures occur precisely where the non-Gaussian structure of the invariant measure matters most: the $\alpha_3$ and $\alpha_4$ perturbations reshape the potential wells and tails in ways that a Gaussian approximation cannot capture. This comparison suggests that quasi-Gaussian closures may be adequate for bulk statistics of low order in weakly non-Gaussian regimes, but unreliable for strongly non-Gaussian or tail-sensitive quantities, motivating the data-driven score estimation methods of Sec.~\ref{subsec:score_estimation}.

\section{\label{sec:o_results} Calibration Results}

We now turn from the local Jacobian comparisons of Sec.~\ref{sec:j_results} to full parameter calibration, where the GFDT-derived sensitivity matrix is embedded in an iterative optimization loop that drives model statistics toward empirical targets. The three case studies collected in this section share the same statistics-matching objective and the same regularized Gauss--Newton backbone, but they represent distinct scientific settings: direct parameter recovery in a scalar stochastic climate model, calibration of a slow--fast system with multiplicative noise, and stochastic closure fitting for unresolved forcing in the two-scale Lorenz--96 equations. This organization highlights the common response-theoretic mechanism across the examples while also showing how the score representation must adapt to dimension, non-Gaussianity, and spatial symmetry.

\medskip
\noindent\textbf{Common calibration protocol.}
All three subsections follow the same outer loop, which we describe here to avoid repetition. Let $\bm{\mathcal{A}}(\bm{x})=(\mathcal{A}_1,\ldots,\mathcal{A}_q)^\top$ denote the vector of observables and $\bm{A}$ their target values (estimated from data or a reference simulation). At each iteration $k$, the algorithm performs three steps: (i) simulate the model at the current parameters $\bm{\theta}^{(k)}$, (ii) estimate the sensitivity matrix from that simulation using the GFDT, and (iii) update the parameters via a regularized Gauss--Newton correction.

More precisely, given a trajectory $\{\bm{x}_t^{(k)}\}_{t=1}^{T}$ of $T$ time steps generated at $\bm{\theta}^{(k)}$, we estimate the current statistic vector by a time average and form the residual:
\begin{equation}
\bm{r}^{(k)} \;=\; \bm{G}^{(k)}-\bm{A},
\qquad
\bm{G}^{(k)} \equiv \frac{1}{T}\sum_{t=1}^{T}\bm{\mathcal{A}}(\bm{x}_t^{(k)}).
\label{eq:residual}
\end{equation}
Here $\bm{G}^{(k)}$ is the empirical (finite-sample) estimator of the exact forward map $\bm{\mathcal{G}}(\bm{\theta}^{(k)})$ defined in Sec.~\ref{sec:Motivation}; ergodicity ensures that $\bm{G}^{(k)} \to \bm{\mathcal{G}}(\bm{\theta}^{(k)})$ as $T \to \infty$. In all three models, one long trajectory serves double duty: it is used both to estimate $\bm{G}^{(k)}$ and to evaluate the GFDT correlation integrals.

The sensitivity matrix at iteration $k$,
\begin{equation}
\bm{S}^{(k)} \;=\; \frac{\partial \langle \bm{\mathcal{A}} \rangle}{\partial \bm{\theta}}\Big|_{\bm{\theta}^{(k)}} \in \mathbb{R}^{q\times p},
\label{eq:sensitivity}
\end{equation}
is estimated from the unperturbed dynamics through the GFDT formulas of Sec.~\ref{subsec:gfdt_jacobian}. By stationarity, each entry admits the two-time correlation representation
\begin{equation}
S_{mj}^{(k)}
=
\int_0^\infty
\Big\langle
\mathcal{A}_m(\bm{x}_t^{(k)})\,
g_{\theta_j}^{(k)}(\bm{x}_{t-s}^{(k)})
\Big\rangle
\,ds,
\label{eq:results_conjugate_form}
\end{equation}
where $g_{\theta_j}$ is the GFDT conjugate observable associated with parameter $\theta_j$. This is the quantity denoted $\mathcal{B}$ in the general GFDT (Eq.~\eqref{eq:gfdt_B_def}), specialized to a step perturbation of parameter $\theta_j$; it encodes the divergence and score-coupling structure derived in Sec.~\ref{subsec:gfdt_jacobian}. The next three subsections make these conjugate observables explicit for each model.

With $\bm{S}^{(k)}$ and $\bm{r}^{(k)}$ in hand, we compute a regularized Gauss--Newton correction:
\begin{equation}
\left((\bm{S}^{(k)})^{\!\top}\bm{B}^{(k)}\bm{S}^{(k)}+\bm{\Gamma}^{(k)}\right)\delta\bm{\theta}^{(k)}
\;=\;-(\bm{S}^{(k)})^{\!\top}\bm{B}^{(k)}\,\bm{r}^{(k)},
\qquad
\bm{\theta}^{(k+1)} \;=\; \bm{\theta}^{(k)} + \delta\bm{\theta}^{(k)}.
\label{eq:GN-step}
\end{equation}
Here $\bm{B}^{(k)} \in \mathbb{R}^{q \times q}$ is the weight matrix for the observables and $\bm{\Gamma}^{(k)} \succeq 0$ is the regularization matrix. Both are updated at each iteration: $\bm{B}^{(k)}$ is recomputed from the empirical covariance of the observables along the current trajectory, and $\bm{\Gamma}^{(k)}$ is adjusted based on conditioning diagnostics. In the scalar and triad examples, $\bm{B}^{(k)}$ is the inverse empirical covariance regularized by a fixed ridge $10^{-10}\bm{I}$, and the iterations are terminated when the mismatch norm $\|\bm{r}^{(k)}\|$ falls below $10^{-3}$ or when the number of iterations exceeds $10$ (see App.~\ref{app:scalar_details} and App.~\ref{app:triad_details} for full details). In the Lorenz--96 experiment, the same Gauss--Newton structure is retained but supplemented by diagonal inverse-variance weighting, column equilibration, and stability safeguards, as detailed in App.~\ref{app:l96_details}; all three Jacobian constructions are advanced for $15$ iterations.

The score representation is adapted to the geometry of each problem. In the scalar and triad models we use KGMM as the non-Gaussian estimator, finite differences as a reference, and a quasi-Gaussian closure as a baseline; for the scalar model, the exact analytic score provides an additional benchmark. In the Lorenz--96 closure problem the state dimension is $36$ and the slow field lives on a periodic lattice. A periodic convolutional U-Net, whose translation-equivariant architecture exploits the lattice symmetry, replaces KGMM as the non-Gaussian score estimator.

\subsection{Application to a Scalar Stochastic Model for Low-Frequency Variability}
\label{sec:reduced1d}

We first apply the calibration framework to a one-dimensional stochastic model for low-frequency climate variability, originally derived by \cite{NormalForms} using stochastic reduction techniques developed in \cite{MTV1,MTV2}. The model represents the motion of an overdamped particle in a quartic potential and captures the asymmetric, weakly non-Gaussian variability typical of low-frequency climate indices. In Langevin notation, the state $x(t)$ evolves according to
\begin{equation}
\dot{x} = F_0 + a x + b x^2 - c x^3 + \sigma\,\xi(t),
\end{equation}
where $\xi$ is a delta-correlated Gaussian white noise. Written in It\^o form, this corresponds to $dx = (F_0 + ax + bx^2 - cx^3)\,dt + \sigma\,dW_t$, so the drift function is a cubic polynomial and the diffusion is a constant. We denote the constant forcing by $F_0$ (rather than $F$) to distinguish it from the general drift function $\bm{F}$ of Sec.~\ref{sec:Motivation}. The parameter vector is $\bm{\theta}=(F_0, a, b, c, \sigma)$, comprising four drift parameters and one diffusion parameter.

The true parameter values used to generate the reference trajectory are
\begin{equation}
F_0=0.6,\quad a=-0.0222,\quad b=-0.2,\quad c=0.0494,\quad \sigma=0.7071.
\end{equation}
The negative quadratic coefficient $b$ introduces skewness, while the positive cubic coefficient $c$ provides saturation at large amplitudes.

Because the noise is additive (the diffusion does not depend on the state), the stationary Fokker--Planck equation can be integrated in closed form. The invariant density is
\begin{equation}
\rho(x) \propto \exp\left[ \frac{2}{\sigma^{2}}\Big(F_0 x + \frac{a}{2}x^2 + \frac{b}{3}x^3 - \frac{c}{4}x^4\Big)\right],
\end{equation}
and the corresponding score function is proportional to the drift divided by the diffusion coefficient:
\begin{equation}
s(x) = \partial_x \log \rho(x) = \frac{2}{\sigma^{2}}\,\Big( F_0 + a x + b x^2 - c x^3\Big).
\end{equation}
This identity---that the score is $2f(x)/\sigma^2$ for a 1D additive-noise SDE with drift $f$---is a special property of one-dimensional systems that will not hold in higher dimensions.

\medskip
\noindent\textbf{GFDT conjugate observables.}
To apply the GFDT, we need the conjugate observable $g_{\theta_j}$ for each parameter $\theta_j$, obtained by specializing the general formulas~\eqref{eq:stat_jac_alpha_limit}--\eqref{eq:stat_jac_beta_limit} to this model. For any scalar observable $\mathcal{A}$, the sensitivity takes the correlation form
\begin{equation}
\frac{\partial\langle \mathcal{A}\rangle}{\partial \theta_j}
=
\int_0^\infty
\big\langle \mathcal{A}(x_t)\,g_{\theta_j}(x_{t-s})\big\rangle\,ds,
\qquad
\theta_j\in\{F_0,a,b,c,\sigma\}.
\label{eq:scalar_gfdt_form}
\end{equation}
The drift parameters enter through the divergence-plus-score structure of~\eqref{eq:stat_jac_alpha_limit}. For example, the parameter Jacobian of the drift with respect to $F_0$ is $\partial_{F_0}(F_0 + ax + bx^2 - cx^3) = 1$ (a constant), so its divergence vanishes and only the score coupling remains, giving $g_{F_0}(x) = -s(x)$. Repeating this calculation for each parameter yields the full set of conjugate observables:
\begin{align}
g_{F_0}(x) &= -s(x), &
g_a(x) &= -1-x\,s(x), &
g_b(x) &= -2x-x^2 s(x), \nonumber\\
g_c(x) &= 3x^2+x^3 s(x), &
g_\sigma(x) &= \sigma\big(s'(x)+s(x)^2\big).
\label{eq:scalar_gfdt_conjugates}
\end{align}
The first four conjugate observables (for the drift parameters) depend only on the score and polynomial functions of $x$. The last one, $g_\sigma$, is the diffusion-response conjugate: it involves the score derivative $s'(x)$ and the squared score $s(x)^2$, arising from the general diffusion-response formula~\eqref{eq:stat_jac_beta_limit} specialized to constant (state-independent) noise. Although the reported calibration updates only the identifiable subset $(F_0,a,\sigma)$, writing the full family clarifies the sensitivity structure summarized in Table~\ref{tab:S_reduced1d}.

\medskip
\noindent\textbf{Observables and calibration setup.}
The observables used for calibration are the mean, the second moment, and a lower-tail exceedance indicator:
\begin{equation}
\bm{\mathcal{A}}(x) = \big(x,\; x^2,\; \mathbf{1}_{x\le x_{\mathrm{th}}} \big),
\end{equation}
where $x_{\mathrm{th}}=2.5$ is a fixed threshold. The three observables probe different aspects of the invariant distribution: the first moment locates the center, the second captures the spread, and the indicator measures the probability mass in the lower tail. We tune the subset $(F_0,a,\sigma)$ while holding $(b,c)$ fixed. This restriction is motivated by identifiability: the parameters $b$ (skewness control) and $c$ (kurtosis/saturation control) affect the statistics in ways that are nearly degenerate with $(F_0,a,\sigma)$ when only three observables are used, so attempting to tune all five simultaneously would lead to an ill-conditioned inverse problem.

The calibration starts from the initial guess
\begin{equation}
F_0^{(0)}=0.72,\quad a^{(0)}=-0.02664,\quad b^{(0)}=-0.20,\quad c^{(0)}=0.0494,\quad \sigma^{(0)}=0.84852,
\end{equation}
which is deliberately biased away from the true values (by $20\%$ in $F_0$, $a$, and $\sigma$). For each calibration iteration, the system is integrated at the current parameter values to generate a trajectory of $T = 10^7$ time steps at $\Delta t = 0.01$, corresponding to a total simulation time of $10^5$ in model units. The decorrelation time of the process is approximately $1$ model time unit, so the trajectory contains roughly $10^5$ effectively independent samples. Full numerical details are given in App.~\ref{app:scalar_details}.

\begin{figure*}[t]
  \centering
  \includegraphics[width=0.95\textwidth]{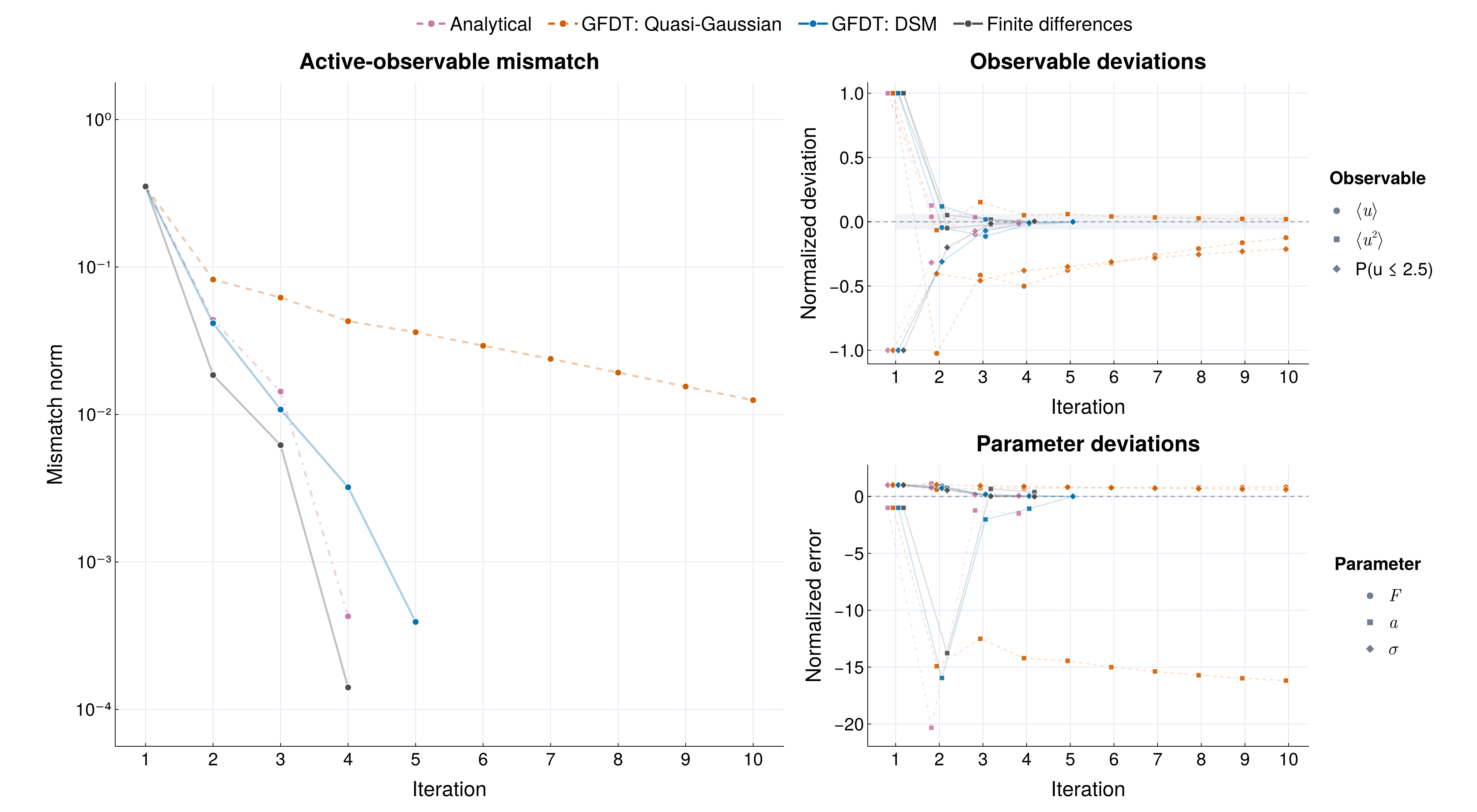}
  \caption[Reduced 1D calibration.]{\textbf{Reduced 1D Calibration.} Left panel: active-observable mismatch norm on a logarithmic scale versus iteration. Top-right panel: normalized observable deviations for $\mathcal{A}_i\in\{\langle x\rangle,\,\langle x^2\rangle,\,P(x\le x_{\mathrm{th}})\}$, each rescaled so that the initial deviation equals $\pm 1$. Bottom-right panel: normalized parameter errors for the tuned parameters $(F_0,a,\sigma)$. Colored curves correspond to the Jacobian surrogate used inside the Gauss--Newton step: Analytical (pink), Finite differences (grey), GFDT: Quasi-Gaussian (orange), and GFDT: DSM (blue). The DSM and analytical Jacobians yield rapid, accurate convergence and agree with the finite-difference baseline, while the quasi-Gaussian closure exhibits systematic bias and substantially slower mismatch reduction.}
  \label{fig:reduced1d_calib}
\end{figure*}

\medskip
\noindent\textbf{Results.}
Figure~\ref{fig:reduced1d_calib} shows the convergence of the calibration. The left panel tracks the overall mismatch norm $\|\bm{G}^{(k)}-\bm{A}\|$ on a logarithmic scale, the top-right panel shows the normalized deviations of each observable from its target, and the bottom-right panel shows the normalized parameter errors. In each panel, all deviations are rescaled by their initial values so that convergence to zero indicates recovery of the target statistics or true parameters. Four variants of the sensitivity matrix are compared: one using the exact analytic score (available for this model), one using DSM-estimated score (via KGMM), one using a quasi-Gaussian closure (which replaces $\rho$ with a Gaussian having the same mean and variance), and one using finite differences as a reference.

The DSM-based GFDT Jacobians are virtually indistinguishable from the analytical ones and closely match the finite-difference baseline, driving the mismatch norm and normalized deviations rapidly toward zero within a few iterations. The quasi-Gaussian closure, by contrast, shows noticeable systematic biases: its mismatch norm decreases much more slowly, and the normalized deviations of the indicator observable and the variance remain far from zero, reflecting the importance of the non-Gaussian tails of the invariant distribution.

Table~\ref{tab:S_reduced1d} reports the full sensitivity matrices computed at the initial parameter guess using the four approaches. Each row corresponds to an observable and each column to a parameter direction, so the table displays $\partial\langle \mathcal{A}_i\rangle / \partial \theta_j$ for all combinations. The DSM (KGMM) estimates closely align with the analytical Jacobian across all entries, confirming that data-driven score estimation yields physically consistent parameter sensitivities even for non-Gaussian observables. The quasi-Gaussian closure departs most visibly in the columns associated with $b$ and $c$ (the nonlinear drift parameters), where it can produce the wrong sign, and in the $\sigma$ column, where the diffusion-response formula amplifies score-approximation errors through the $s'(x) + s(x)^2$ structure.

\begin{table*}[t]
\centering
\footnotesize
\caption[Parameter Jacobians for the scalar model.]{Parameter Jacobians $S=\partial\langle\bm{\mathcal{A}}\rangle/\partial\bm{\theta}$ for the scalar model. Rows correspond to $\mathcal{A}_1=x$, $\mathcal{A}_2=x^2$, $\mathcal{A}_3=\mathbf{1}_{x\le x_{\mathrm{th}}}$ with $x_{\mathrm{th}}=2.5$. Columns correspond to $(F_0, a, b, c, \sigma)$.}
\label{tab:S_reduced1d}
\begin{tabular}{lrrrrr}
\toprule
Method & $\partial/\partial F_0$ & $\partial/\partial a$ & $\partial/\partial b$ & $\partial/\partial c$ & $\partial/\partial \sigma$ \\
\midrule
\multicolumn{6}{l}{Analytical} \\
\midrule
$x$              &  1.523101e+00 &  1.451248e+00 &  2.743243e+00 & -3.926157e+00 & -7.433430e-01 \\
$x^2$            &  2.903625e+00 &  4.441893e+00 &  6.756236e+00 & -1.419935e+01 &  1.895477e-01 \\
$\mathbf{1}_{x\le x_{\mathrm{th}}}$ & -1.067605e-01 & -1.993064e-01 & -4.237086e-01 &  9.391258e-01 & -1.404245e-01 \\
\midrule
\multicolumn{6}{l}{GFDT: DSM} \\
\midrule
$x$              &  1.580722e+00 &  1.379337e+00 &  2.985298e+00 & -3.503986e+00 & -8.312558e-01 \\
$x^2$            &  2.927969e+00&  4.531903e+00 &  6.709038e+00 & -1.460211e+01 &  1.945329e-01 \\
$\mathbf{1}_{x\le x_{\mathrm{th}}}$ & -1.081247e-01 & -1.992222e-01 & -4.265260e-01 &  9.342348e-01 & -1.394241e-01 \\
\midrule
\multicolumn{6}{l}{GFDT: Quasi-Gaussian} \\
\midrule
$x$               &  2.129722e+00 & -7.574723e-01 &  5.434414e+00 &  1.156984e+01 & -5.668083e+00 \\
$x^2$            &  2.087184e+00 &  3.738908e+00 & -4.596963e-03 & -1.536090e+01 &  1.425404e+00 \\
$\mathbf{1}_{x\le x_{\mathrm{th}}}$ & -7.933504e-02 & -9.021017e-02 & -2.319406e-01 &  2.298429e-01 &  2.667446e-02 \\
\midrule
\multicolumn{6}{l}{Finite differences} \\
\midrule
$x$               &  1.500011e+00 &  1.355209e+00 &  2.799968e+00 &  -3.416764e+00 & -7.361646e-01 \\
$x^2$            &  2.718941e+00 &  4.258402e+00 &  6.260473e+00 &  -1.374833e+01 &  2.148115e-01 \\
$\mathbf{1}_{x\le x_{\mathrm{th}}}$ & -8.999100e-02 & -1.699830e-01 &  -3.749625e-01 &  8.399160e-01 & -1.149885e-01 \\
\bottomrule
\end{tabular}
\end{table*}

\subsection{Slow--Fast Triad Model and Application to ENSO}
\label{sec:triad_enso}

We next consider a three-dimensional slow--fast triad model introduced by \cite{thual2016simple} as a conceptual representation of the El Ni\~no--Southern Oscillation (ENSO). The model couples two slow variables $(u_1,u_2)$, representing ocean--atmosphere modes, to a fast variable $\tau$, representing wind-burst activity. The slow variables form a damped linear oscillator with frequency $\omega$ and common damping rate $d_u$, while the fast variable relaxes toward zero with rate $d_{\tau}$ and injects energy into the first slow mode through a linear coupling. A key physical feature is that the noise driving $\tau$ is multiplicative: its amplitude grows with $u_1$ through the function $\tanh(u_1)+1$, capturing the observation that westerly wind bursts are more active during warm ENSO states. In physical variables $(u_1,u_2,\tau)$ the dynamics read
\begin{equation}
\begin{aligned}
\dot{u}_1 &= -d_{u}\,u_1 - \omega\,u_2 + \tau + \sigma_1\,\xi_1(t),\\
\dot{u}_2 &= -d_{u}\,u_2 + \omega\,u_1 + \sigma_2\,\xi_2(t),\\
\dot{\tau} &= -d_{\tau}\,\tau + \sigma_3\,\big(\tanh(u_1)+1\big)\,\xi_3(t),
\end{aligned}
\end{equation}
where $\xi_1, \xi_2, \xi_3$ are independent Gaussian white noises. The parameter vector is $\bm{\theta}=(d_u,\omega,d_{\tau},\sigma_1,\sigma_2,\sigma_3)$: three drift parameters governing the deterministic dynamics and three diffusion parameters controlling the noise amplitudes. The noise in the first two equations is additive, while the noise in the third equation is state-dependent (multiplicative) through the factor $\tanh(u_1)+1$.

\medskip
\noindent\textbf{Observables.}
The observables used for calibration are the six distinct second-order moments and cross-moments:
\begin{equation}
\bm{\mathcal{A}}(u_1,u_2,\tau)=\big(u_1^2,\;u_2^2,\;u_1u_2,\;\tau^2,\;u_1\tau,\;u_2\tau\big).
\end{equation}
These probe the variance of each variable, the coupling between the slow oscillator modes, and the covariance between the fast wind-burst variable and the slow ocean--atmosphere modes. Together they characterize the second-order structure of the coupled variability.

\medskip
\noindent\textbf{True parameters and initial guess.}
The true parameters are
\begin{equation}
d_u=0.2,\quad \omega=0.4,\quad d_{\tau}=2.0,\quad \sigma_1=0.3,\quad \sigma_2=0.3,\quad \sigma_3=1.5,
\end{equation}
and the calibration starts from the biased initial guess
\begin{equation}
d_u^{(0)}=0.24,\; \omega^{(0)}=0.32,\; d_{\tau}^{(0)}=2.2,\; \sigma_1^{(0)}=0.36,\; \sigma_2^{(0)}=0.27,\; \sigma_3^{(0)}=1.575.
\end{equation}

\medskip
\noindent\textbf{GFDT conjugate observables.}
The triad model presents additional challenges relative to the scalar case because of the multiplicative noise in the $\tau$ equation. The GFDT framework handles this through the diffusion-response formula~\eqref{eq:stat_jac_beta_limit}, which introduces score derivatives into the conjugate observables for the noise parameters. Writing $\bm{u}=(u_1,u_2,\tau)^\top$, $q(u_1)=1+\tanh(u_1)$ for the state-dependent noise modulation, and
\[
\bm{s}(\bm{u})=(s_1(\bm{u}),s_2(\bm{u}),s_3(\bm{u}))^\top=\nabla_{\bm{u}}\log\rho(\bm{u})
\]
for the three-dimensional score of the invariant density, each sensitivity entry again takes the GFDT correlation form:
\begin{equation}
\frac{\partial\langle \mathcal{A}\rangle}{\partial \theta_j}
=
\int_0^\infty
\big\langle \mathcal{A}(\bm{u}_t)\,g_{\theta_j}(\bm{u}_{t-s})\big\rangle\,ds,
\qquad
\theta_j\in\{d_u,\omega,d_\tau,\sigma_1,\sigma_2,\sigma_3\}.
\label{eq:triad_gfdt_form}
\end{equation}
The conjugate observables for the three drift parameters involve only the score:
\begin{align}
g_{d_u}(\bm{u}) &= 2+u_1 s_1(\bm{u})+u_2 s_2(\bm{u}), &
g_{\omega}(\bm{u}) &= u_2 s_1(\bm{u})-u_1 s_2(\bm{u}), &
g_{d_\tau}(\bm{u}) &= 1+\tau s_3(\bm{u}).
\label{eq:triad_drift_conjugates}
\end{align}
These follow from~\eqref{eq:stat_jac_alpha_limit} by computing the divergence and score product for each drift parameter's Jacobian. For instance, $g_{d_u}$ arises because $\partial_{d_u}$ of the drift vector $(-d_u u_1, -d_u u_2, 0)^\top$ is $(-u_1, -u_2, 0)^\top$, whose divergence is $-2$ and whose inner product with the score is $-u_1 s_1 - u_2 s_2$.

The conjugate observables for the noise parameters involve score derivatives, reflecting the diffusion-response structure of~\eqref{eq:stat_jac_beta_limit}:
\begin{align}
g_{\sigma_1}(\bm{u}) &= \sigma_1\big(\partial_{u_1}s_1(\bm{u})+s_1(\bm{u})^2\big), &
g_{\sigma_2}(\bm{u}) &= \sigma_2\big(\partial_{u_2}s_2(\bm{u})+s_2(\bm{u})^2\big), &
g_{\sigma_3}(\bm{u}) &= \sigma_3 q(u_1)^2\big(\partial_{\tau}s_3(\bm{u})+s_3(\bm{u})^2\big).
\label{eq:triad_diff_conjugates}
\end{align}
The first two ($g_{\sigma_1}$ and $g_{\sigma_2}$) have the same form as $g_\sigma$ in the scalar model, because the noise in the $u_1$ and $u_2$ equations is additive. The third, $g_{\sigma_3}$, additionally carries the factor $q(u_1)^2 = (1+\tanh(u_1))^2$, which arises because the diffusion matrix element $\sigma_3 q(u_1)$ is state-dependent: the chain rule through~\eqref{eq:Ktilde_def} produces this squared modulation. Estimating all six conjugate observables requires not only the score $\bm{s}$ but also the diagonal entries of the score Jacobian ($\partial_{u_1}s_1$, $\partial_{u_2}s_2$, $\partial_{\tau}s_3$), which we obtain by automatic differentiation of the KGMM interpolant.

\medskip
\noindent\textbf{Simulation and calibration.}
For each calibration iteration, the system is integrated at the current parameter values to generate a trajectory of $T = 10^7$ time steps at $\Delta t = 0.01$. The decorrelation time of the slow oscillator is approximately $1/d_u = 5$ model time units, so the trajectory contains roughly $2 \times 10^4$ effectively independent slow-mode samples. Because the stationary density is not available in closed form for this three-dimensional system, we compare three Jacobian constructions: (i) DSM-based score estimates (via KGMM), (ii) a quasi-Gaussian closure, and (iii) a finite-difference baseline computed by re-simulating at perturbed parameter values. Full numerical details are given in App.~\ref{app:triad_details}.

\begin{figure*}[t]
  \centering
  \includegraphics[width=0.98\textwidth]{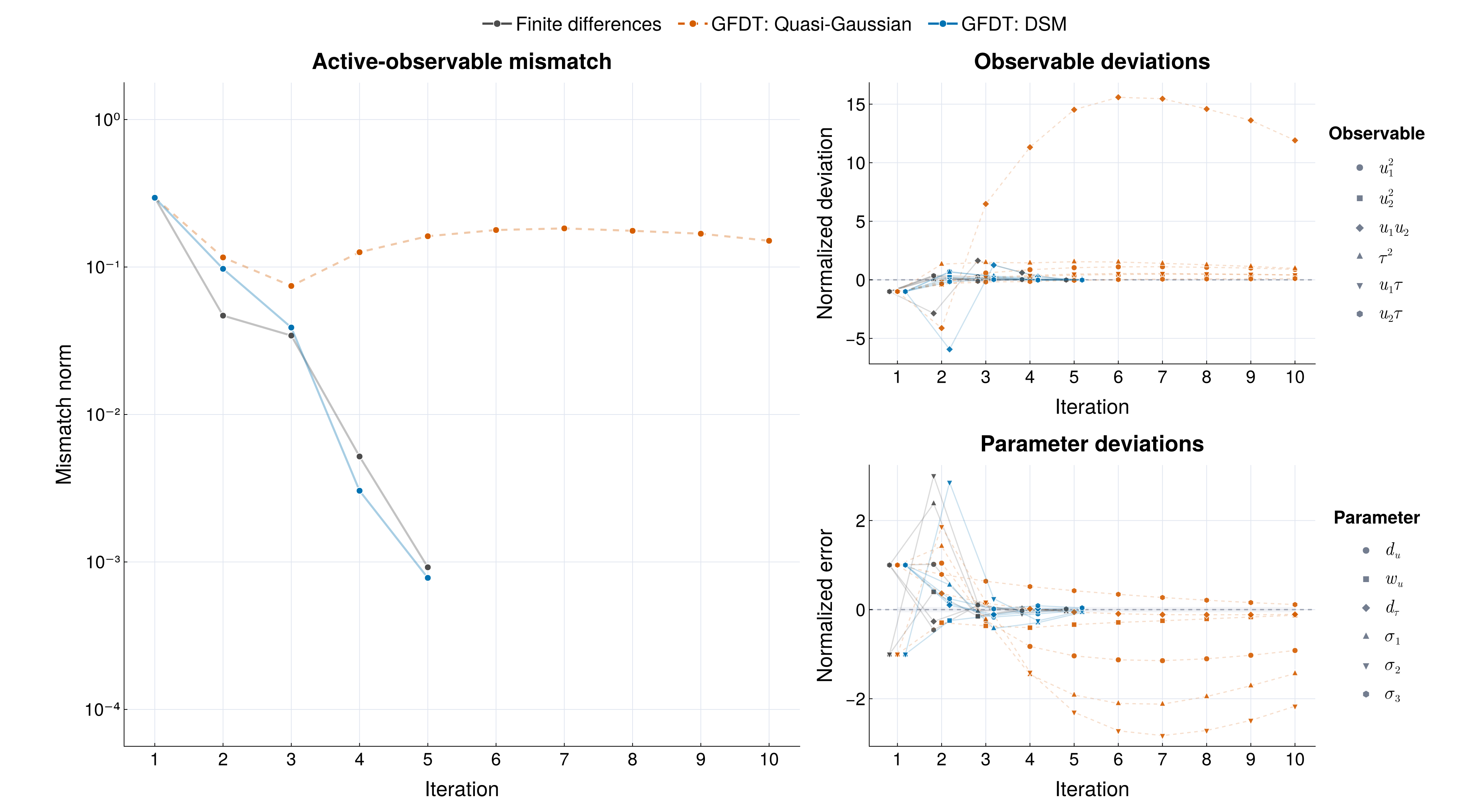}
  \caption[ENSO calibration.]{\textbf{ENSO Calibration.} Left panel: active-observable mismatch norm on a logarithmic scale versus iteration. Top-right panel: normalized observable deviations for the six second-order observables $\{u_1^2,\,u_2^2,\,u_1u_2,\,\tau^2,\,u_1\tau,\,u_2\tau\}$, each rescaled so that the initial deviation equals $\pm 1$. Bottom-right panel: normalized parameter errors for $(d_u,\omega,d_\tau,\sigma_1,\sigma_2,\sigma_3)$. Colored curves: Finite differences (grey), GFDT: Quasi-Gaussian (orange), and GFDT: DSM (blue). The DSM-based Jacobians steer the iteration rapidly to the correct parameters and statistics, whereas the quasi-Gaussian closure shows systematic errors, especially in the mismatch norm and the $\tau$-related moments.}
  \label{fig:triad_calib}
\end{figure*}

\medskip
\noindent\textbf{Results.}
Figure~\ref{fig:triad_calib} shows the calibration trajectories. The left panel tracks the mismatch norm, and the right panels show the normalized observable deviations and parameter errors. The DSM-based Jacobians steer the iteration to the correct parameters and target statistics, closely tracking the finite-difference reference: both the mismatch norm and the normalized deviations converge rapidly to zero. The quasi-Gaussian closure shows systematic errors, particularly in the $\tau$-related moments ($\tau^2$, $u_1\tau$, $u_2\tau$), where the multiplicative noise generates non-Gaussian correlations that the closure cannot capture.

Table~\ref{tab:S_triad} compiles the sensitivity matrices computed at the initial parameter guess. Each row is an observable, each column a parameter direction, and each block corresponds to one of the three Jacobian methods. Agreement between DSM (KGMM) and finite differences is excellent across all entries. The quasi-Gaussian closure departs markedly in the columns associated with $\sigma_3$ and $d_{\tau}$, which are the parameters most directly coupled to the non-Gaussian, multiplicative-noise structure of the $\tau$ equation.

\begin{table*}[t]
\centering
\scriptsize
\caption[Parameter Jacobians for the triad/ENSO model.]{Parameter Jacobians $S=\partial\langle\bm{\mathcal{A}}\rangle/\partial\bm{\theta}$ for the triad/ENSO model. Rows are $\{u_1^2, u_2^2, u_1u_2, \tau^2, u_1\tau, u_2\tau\}$ and columns are $(d_u,\omega,d_\tau,\sigma_1,\sigma_2,\sigma_3)$.}
\label{tab:S_triad}
\begin{tabular}{lrrrrrr}
\toprule
Method & $\partial/\partial d_u$ & $\partial/\partial \omega$ & $\partial/\partial d_{\tau}$ & $\partial/\partial \sigma_1$ & $\partial/\partial \sigma_2$ & $\partial/\partial \sigma_3$ \\
\midrule
\multicolumn{7}{l}{Finite differences} \\
\midrule
$u_1^2$  & -3.050743e+00 & -7.758489e-01 & -5.495300e-01 &  1.209837e+00 &  3.740469e-01 &  8.258673e-01 \\
$u_2^2$  & -3.508226e+00 &  9.182873e-01 & -3.071658e-01 &  5.772242e-01 &  7.840313e-01 &  4.317518e-01 \\
$u_1u_2$ & -9.582175e-01 & -2.166181e-01 & -2.328459e-01 &  4.310720e-01 & -2.615223e-01 &  3.277357e-01 \\
$\tau^2$  & -1.889321e-01 &  1.442077e-01 & -2.897416e-01 &  1.758095e-01 &  3.114253e-02 &  8.329819e-01 \\
$u_1\tau$ & -1.509486e-01 &  3.113702e-02 & -2.112176e-01 &  7.027646e-02 &  7.297418e-03 &  3.102760e-01 \\
$u_2\tau$ & -3.162352e-02 &  8.415626e-02 & -3.774696e-02 &  5.875199e-03 &  8.329266e-04 &  3.670596e-02 \\
\midrule
\multicolumn{7}{l}{GFDT: DSM} \\
\midrule
$u_1^2$  & -2.924411e+00 & -6.936217e-01 & -4.124085e-01 &  1.314130e+00 &  3.713429e-01 &  6.036389e-01 \\
$u_2^2$  & -3.278611e+00 &  1.044633e+00 & -2.411809e-01 &  6.451183e-01 &  7.838778e-01 &  3.216192e-01 \\
$u_1u_2$ & -8.664064e-01 & -1.585690e-01 & -1.725149e-01 &  4.537153e-01 & -2.237340e-01 &  1.815809e-01 \\
$\tau^2$  & -1.598740e-01 &  1.216938e-01 & -3.872378e-01 &  2.438810e-01 &  2.805538e-02 &  1.099367e+00 \\
$u_1\tau$ & -1.612445e-01 &  5.875598e-02 & -2.077249e-01 &  1.357561e-01 &  8.127335e-03 &  2.464953e-01 \\
$u_2\tau$ & -3.201668e-02 &  1.272658e-01 & -3.610327e-02 &  3.027318e-02 &  8.037050e-03 & -2.737315e-03 \\
\midrule
\multicolumn{7}{l}{GFDT: Quasi-Gaussian} \\
\midrule
$u_1^2$  & -5.499362e+00 & -1.031043e+00 & -2.986248e+00 &  1.722336e+00 &  5.188293e-01 &  1.879337e+01 \\
$u_2^2$  & -7.094832e+00 &  1.476745e+00 & -2.386642e+00 &  1.162480e+00 &  1.441148e+00 &  1.623342e+01 \\
$u_1u_2$ & -2.027037e+00 & -2.559634e-01 & -1.362706e+00 &  6.636498e-01 & -3.406273e-01 &  8.941124e+00 \\
$\tau^2$  & -1.029504e-01 & -1.960733e-01 & -1.367300e+00 &  1.184367e-01 & -2.583208e-01 &  8.076022e+00 \\
$u_1\tau$ & -2.810736e-01 & -5.833484e-02 & -8.018629e-01 &  5.513819e-02 & -8.125723e-02 &  4.012663e+00 \\
$u_2\tau$ & -5.586091e-02 &  1.502120e-01 & -1.168387e-01 & -6.854179e-03 & -1.862000e-02 &  3.802178e-01 \\
\bottomrule
\end{tabular}
\end{table*}

Together, these first two case studies demonstrate that the GFDT, when paired with an accurate score estimator such as KGMM, provides reliable parameter Jacobians for both additive and multiplicative noise systems. The close agreement with finite differences in both models, together with the successful recovery of the true parameters from biased initial guesses, validates the end-to-end calibration approach. We now turn to a substantially more challenging setting where the calibration target is statistical fidelity of a reduced closure, rather than parameter identification.

\subsection{Stochastic Closure Calibration in the Lorenz--96 System}
\label{sec:l96}

The final calibration problem is the most demanding: fitting a stochastic closure for the two-scale Lorenz--96 system \cite{lorenz1996predictability}, a canonical multiscale testbed for unresolved fast processes and stochastic parameterization. Unlike the previous two examples, where the target statistics were generated by the same model family used in calibration (so exact parameter recovery is in principle achievable), here the reduced model is only an approximate closure ansatz for a genuinely multiscale truth. The calibration objective is therefore statistical fidelity rather than parameter identification: we seek closure coefficients such that the invariant law of the reduced dynamics reproduces key equilibrium statistics of the resolved variables in the full two-scale system.

\medskip
\noindent\textbf{The two-scale Lorenz--96 system.}
The full model couples $K$ slow (resolved) variables $\{X_k\}$ to $K \times J$ fast (unresolved) variables $\{Y_{j,k}\}$:
\begin{align}
dX_k &= \left[- X_{k-1}\bigl(X_{k-2}-X_{k+1}\bigr) - X_k + \mathcal{F} - \frac{hc}{J}\sum_{j=1}^J Y_{j,k}\right]dt + \sigma_x\, dW_k^x, \label{eq:l96_X} \\
dY_{j,k} &= \left[-cb\,Y_{j+1,k}\bigl(Y_{j+2,k}-Y_{j-1,k}\bigr) - cY_{j,k} + \frac{hc}{J}X_k\right]dt + \sigma_y\, dW_{j,k}^y, \label{eq:l96_Y}
\end{align}
with indices understood modulo $K$ in $k$ and modulo $J$ in $j$. We denote the external forcing by $\mathcal{F}$ to distinguish it from the drift function $\bm{F}$ of Sec.~\ref{sec:Motivation}. The parameters $h$ and $c$ control the coupling strength and time-scale separation between slow and fast variables, while $b$ sets the amplitude ratio of the fast dynamics. Throughout this subsection we use
\[
K=36,\quad J=10,\quad \mathcal{F}=10,\quad h=1,\quad c=10,\quad b=10,\quad \sigma_x=1,\quad \sigma_y=0,\quad \Delta t = 0.005.
\]

The fast $Y$ variables produce a structured, state-dependent, and temporally correlated forcing on the slow variables through the coupling term $-(hc/J)\sum_j Y_{j,k}$ in~\eqref{eq:l96_X}. The direct noise term $\sigma_x\,dW_k^x$ represents a more rapidly decorrelating remainder associated with unresolved influences that are not captured even by the $Y$ tier. These two sources of randomness play distinct physical roles: the $Y$-mediated contribution carries coherent subgrid backscatter across the resolved dynamics, while the additive white noise accounts for residual small-scale variability that is effectively uncorrelated at the slow time scale.

\medskip
\noindent\textbf{The reduced stochastic closure.}
Following the stochastic parameterization strategy of Arnold et al.~\cite{arnold2013stochastic}, we retain only the slow variables and replace the unresolved coupling by a local cubic polynomial plus additive white noise:
\begin{equation}
dX_k
=
\Big[
-X_{k-1}(X_{k-2}-X_{k+1}) - X_k + \mathcal{F}
- \bigl(\alpha_0+\alpha_1 X_k+\alpha_2 X_k^2+\alpha_3 X_k^3\bigr)
\Big]dt
+ \sigma\, dW_k,
\label{eq:l96_stochastic}
\end{equation}
where the Wiener processes $W_k$ are independent across lattice sites. The closure parameter vector is
\begin{equation}
\bm{\theta} = (\alpha_0,\alpha_1,\alpha_2,\alpha_3,\sigma)^\top \in \mathbb{R}^5,
\label{eq:l96_theta}
\end{equation}
comprising four drift parameters $(\alpha_0,\alpha_1,\alpha_2,\alpha_3)$ that define the polynomial closure and one diffusion parameter $\sigma$. In the notation of Sec.~\ref{sec:Motivation}, $\bm{\alpha}=(\alpha_0,\alpha_1,\alpha_2,\alpha_3)$ and $\bm{\beta}=(\sigma)$. This closure class is intentionally restrictive: it is local in the lattice index $k$, Markovian in time, and acts only on the resolved variables, so there is no reason to expect exact reproduction of the full two-scale statistics.

\medskip
\noindent\textbf{Target observables.}
We calibrate the reduced model against five translation-invariant statistics of the slow field. Writing $\mu(X)=\frac{1}{K}\sum_{k=1}^K X_k$ for the spatial mean, the five observables are:
\begin{equation}
\bm{\Phi}(X)=\bigl(\mu(X),\;V(X),\;\operatorname{Sk}(X),\;\operatorname{Ku}(X),\;C_1(X)\bigr)^\top,
\label{eq:l96_forward_map_impl}
\end{equation}
where
\begin{align}
V(X) &= \frac{1}{K}\sum_{k=1}^K \bigl(X_k-\mu(X)\bigr)^2 \quad &(\text{spatial variance}), \\
\operatorname{Sk}(X) &= \frac{\frac{1}{K}\sum_{k=1}^K \bigl(X_k-\mu(X)\bigr)^3}{V(X)^{3/2}} \quad &(\text{spatial skewness}), \\
\operatorname{Ku}(X) &= \frac{\frac{1}{K}\sum_{k=1}^K \bigl(X_k-\mu(X)\bigr)^4}{V(X)^2}-3 \quad &(\text{spatial excess kurtosis}), \\
C_1(X) &= \frac{1}{K}\sum_{k=1}^K \bigl(X_k-\mu(X)\bigr)\bigl(X_{k+1}-\mu(X)\bigr) \quad &(\text{nearest-neighbor spatial covariance}),
\end{align}
with indices understood modulo $K$. These five observables separately probe the mean level, spread, asymmetry, tail weight, and spatial correlation structure of the slow field. Each is computed as a spatial average over the $K=36$ lattice sites at each time step, exploiting the statistical homogeneity of the periodic Lorenz--96 system to reduce sampling variance. The forward map is then $\bm{\mathcal{G}}(\bm{\theta}) = \langle \bm{\Phi}(X) \rangle_{\bm{\theta}}$, where the expectation is taken over the invariant measure of the reduced model~\eqref{eq:l96_stochastic}.

\medskip
\noindent\textbf{Initialization.}
The target vector $\bm{A}$ and the initial closure guess are both derived from a single long equilibrium trajectory of the two-scale system~\eqref{eq:l96_X}--\eqref{eq:l96_Y}. From this trajectory we diagnose the unresolved tendency acting on each slow variable, fit its conditional mean by a local cubic polynomial in $X_k$, and estimate the residual noise amplitude from the residual autocovariance. This yields the dynamically informed starting guess
\begin{equation}
\bm{\theta}^{(0)}
=
\bigl(
0.1494,\;
0.3945,\;
-0.0084,\;
-0.0017,\;
1.0359
\bigr)^\top,
\label{eq:l96_theta0_main}
\end{equation}
which serves as an initialization for the iterative calibration rather than a final estimate.

\medskip
\noindent\textbf{GFDT conjugate observables.}
For the reduced model~\eqref{eq:l96_stochastic}, the drift perturbation associated with each polynomial coefficient $\alpha_p$ ($p=0,1,2,3$) acts locally on each lattice site:
\[
\bigl(\partial_{\alpha_p} b_{\bm{\theta}}(X)\bigr)_k=-X_k^p,
\]
and the diffusion perturbation associated with $\sigma$ is simply $\partial_\sigma \bm{\Sigma}=\bm{I}_{K}$, the $K$-dimensional identity matrix. Applying the GFDT formulas~\eqref{eq:stat_jac_alpha_limit}--\eqref{eq:stat_jac_beta_limit} yields the conjugate observables
\begin{align}
g_{\alpha_0}(X) &= \sum_{k=1}^K s_k(X), \label{eq:l96_g_alpha0}\\
g_{\alpha_1}(X) &= K + \sum_{k=1}^K X_k\, s_k(X), \\
g_{\alpha_2}(X) &= 2\sum_{k=1}^K X_k + \sum_{k=1}^K X_k^2\, s_k(X), \\
g_{\alpha_3}(X) &= 3\sum_{k=1}^K X_k^2 + \sum_{k=1}^K X_k^3\, s_k(X), \\
g_{\sigma}(X) &= \sigma\Big(\nabla_X\!\cdot\, \bm{s}(X)+\|\bm{s}(X)\|_2^2\Big),
\label{eq:l96_g_sigma}
\end{align}
where $\bm{s}(X)=(s_1(X),\ldots,s_K(X))^\top=\nabla_X\log\rho_{\bm{\theta}}(X)$ is the $K$-dimensional score of the invariant density of the reduced model. The drift conjugates ($g_{\alpha_0}$ through $g_{\alpha_3}$) follow the same divergence-plus-score pattern seen in the scalar and triad models. The diffusion conjugate $g_\sigma$ involves the divergence of the score, $\nabla_X \cdot \bm{s} = \sum_{k=1}^K \partial_{X_k} s_k$, and the squared norm $\|\bm{s}\|_2^2 = \sum_{k=1}^K s_k^2$. In a $K=36$-dimensional state space, evaluating the divergence $\nabla_X \cdot \bm{s}$ naively would require $K$ backward passes through the score network. To avoid this cost, we approximate it using a Hutchinson trace estimator \cite{hutchinson1989stochastic} with $10$ Rademacher probes (see App.~\ref{app:l96_details}).

The sensitivity matrix then takes the explicit correlation form
\begin{equation}
S_{mj}
=
\int_0^\infty
\Big\langle
\Phi_m(X_t)\,
g_{\theta_j}(X_{t-s})
\Big\rangle
\,ds,
\qquad
\theta_j\in\{\alpha_0,\alpha_1,\alpha_2,\alpha_3,\sigma\},
\label{eq:l96_gfdt_form}
\end{equation}
so the Lorenz--96 closure problem exercises both the drift-response and diffusion-response components of the GFDT.

\medskip
\noindent\textbf{Score representation.}
In the scalar and triad models, KGMM was effective because the score could be represented accurately in a low-dimensional state space. Here the state dimension is $K=36$, and the slow field is defined on a periodic lattice with translational symmetry. A periodic convolutional U-Net is therefore a natural choice: it exploits locality (each $s_k$ depends primarily on nearby $X$ values) and translation equivariance (the score function has the same functional form at every lattice site) by construction. A clustering-based KGMM surrogate would ignore this structure and become statistically inefficient in a 36-dimensional space. For comparison, we also compute a quasi-Gaussian GFDT Jacobian (using the empirical Gaussian fitted to the trajectory) and a finite-difference reference Jacobian (requiring $10$ perturbed integrations per iteration for the five-parameter vector).

\medskip
\noindent\textbf{Simulation protocol.}
At each calibration iteration, we integrate the reduced model~\eqref{eq:l96_stochastic} over $[0,5\times10^5]$ with time step $\Delta t=0.005$ and discard the first $6$ time units as transient. The post-transient trajectory is reused for three purposes: estimating the observable vector $\bm{G}^{(k)}$, fitting the U-Net and Gaussian score surrogates from snapshots saved every $200\Delta t=1$, and evaluating the GFDT correlation integrals from a block of $10^6$ samples at cadence $2\Delta t=0.01$. A single unperturbed integration per iteration therefore supplies both GFDT branches. Full numerical details are given in App.~\ref{app:l96_details}.

\medskip
\noindent\textbf{Results: Jacobian comparison.}
Table~\ref{tab:l96_jacobians_run081_iter1} reports the sensitivity matrix at the initial guess~\eqref{eq:l96_theta0_main}. Since the observable vector and parameter vector both have dimension five, the Jacobian is a square $5 \times 5$ matrix. This does not make the problem well-determined in the sense of parameter identification: even with an exact Jacobian and perfect optimization, there is no reason to expect an exact root of $\bm{\mathcal{G}}(\bm{\theta})-\bm{A}=0$ to exist within the restricted closure ansatz~\eqref{eq:l96_stochastic}.

Among the two GFDT variants, the U-Net-based Jacobian is in markedly better agreement with the finite-difference reference than the quasi-Gaussian Jacobian. It reproduces the correct sign pattern and the dominant drift sensitivities across the five observables. Agreement is visibly weaker in the diffusion column $\partial/\partial\sigma$ than in the drift columns, which is expected: the diffusion response depends on $g_\sigma$ and therefore on the divergence $\nabla_X \cdot \bm{s}(X)$, so differentiating the score field introduces additional numerical noise. Even so, the U-Net GFDT Jacobian remains substantially closer to finite differences than the Gaussian GFDT Jacobian and captures the sensitivity structure needed for calibration.

\begin{table*}[t]
\centering
\scriptsize
\caption[Iteration-1 parameter Jacobians for the Lorenz--96 closure problem.]{
Parameter Jacobians $S=\partial\bm{\mathcal{G}}/\partial\bm{\theta}$ at the initial parameter guess \eqref{eq:l96_theta0_main} for the Lorenz--96 closure experiment. Rows are the active observables $(\mu,V,\operatorname{Sk},\operatorname{Ku},C_1)$ and columns are the closure parameters $(\alpha_0,\alpha_1,\alpha_2,\alpha_3,\sigma)$. The three blocks report the numerical finite-difference reference, the GFDT Jacobian based on the periodic U-Net score representation, and the GFDT Jacobian based on the quasi-Gaussian score surrogate.
}
\label{tab:l96_jacobians_run081_iter1}
\begin{tabular}{lrrrrr}
\toprule
Method & $\partial/\partial \alpha_0$ & $\partial/\partial \alpha_1$ & $\partial/\partial \alpha_2$ & $\partial/\partial \alpha_3$ & $\partial/\partial \sigma$ \\
\midrule
\multicolumn{6}{l}{Finite differences} \\
\midrule
$\mu$                & -1.912208e-01 & -4.718669e-01 & -5.578590e+00 & -3.617055e+01 & -1.107000e-02 \\
$V$                  & -2.904590e+00 & -2.024995e+01 & -1.425614e+02 & -1.273573e+03 &  8.771940e-01 \\
$\operatorname{Sk}$  & -2.920575e-02 & -1.599108e-01 & -2.626506e+00 & -2.310829e+01 & -3.238000e-03 \\
$\operatorname{Ku}$  & -6.245075e-02 & -8.355945e-01 & -6.810124e+00 & -8.702567e+01 &  1.176670e-01 \\
$C_1$                & -2.029332e-02 &  5.382703e-01 &  8.567629e+00 &  7.297605e+01 &  1.526460e-01 \\
\midrule
\multicolumn{6}{l}{GFDT: DSM} \\
\midrule
$\mu$                & -1.583310e-01 & -2.423258e-01 & -4.574638e+00 & -2.890482e+01 & -1.688600e-02 \\
$V$                  & -2.736109e+00 & -1.963779e+01 & -1.414838e+02 & -1.243773e+03 &  9.499230e-01 \\
$\operatorname{Sk}$  & -2.288088e-02 & -1.813202e-01 & -2.827964e+00 & -2.562588e+01 & -1.293400e-02 \\
$\operatorname{Ku}$  & -4.247188e-02 & -7.495282e-01 & -6.628061e+00 & -8.357652e+01 &  1.385390e-01 \\
$C_1$                & -7.513799e-02 &  3.167749e-01 &  7.609634e+00 &  7.452266e+01 & -6.859000e-02 \\
\midrule
\multicolumn{6}{l}{GFDT: Quasi-Gaussian} \\
\midrule
$\mu$                & -4.226070e-01 & -1.622865e+00 & -1.186973e+01 & -8.740903e+01 &  8.716400e-02 \\
$V$                  & -3.073120e+00 & -1.322416e+01 & -9.863874e+01 & -8.019176e+02 &  6.917850e-01 \\
$\operatorname{Sk}$  & -3.725698e-02 & -1.217408e-01 & -2.243224e+00 & -1.746171e+01 &  5.290000e-03 \\
$\operatorname{Ku}$  & -1.864844e-01 & -1.049633e+00 & -7.778527e+00 & -8.020019e+01 &  7.887600e-02 \\
$C_1$                &  1.503806e-01 &  5.761944e-01 &  1.025063e+01 &  5.903639e+01 & -1.952700e-02 \\
\bottomrule
\end{tabular}
\end{table*}

\begin{figure*}[t]
  \centering
  \includegraphics[width=\textwidth]{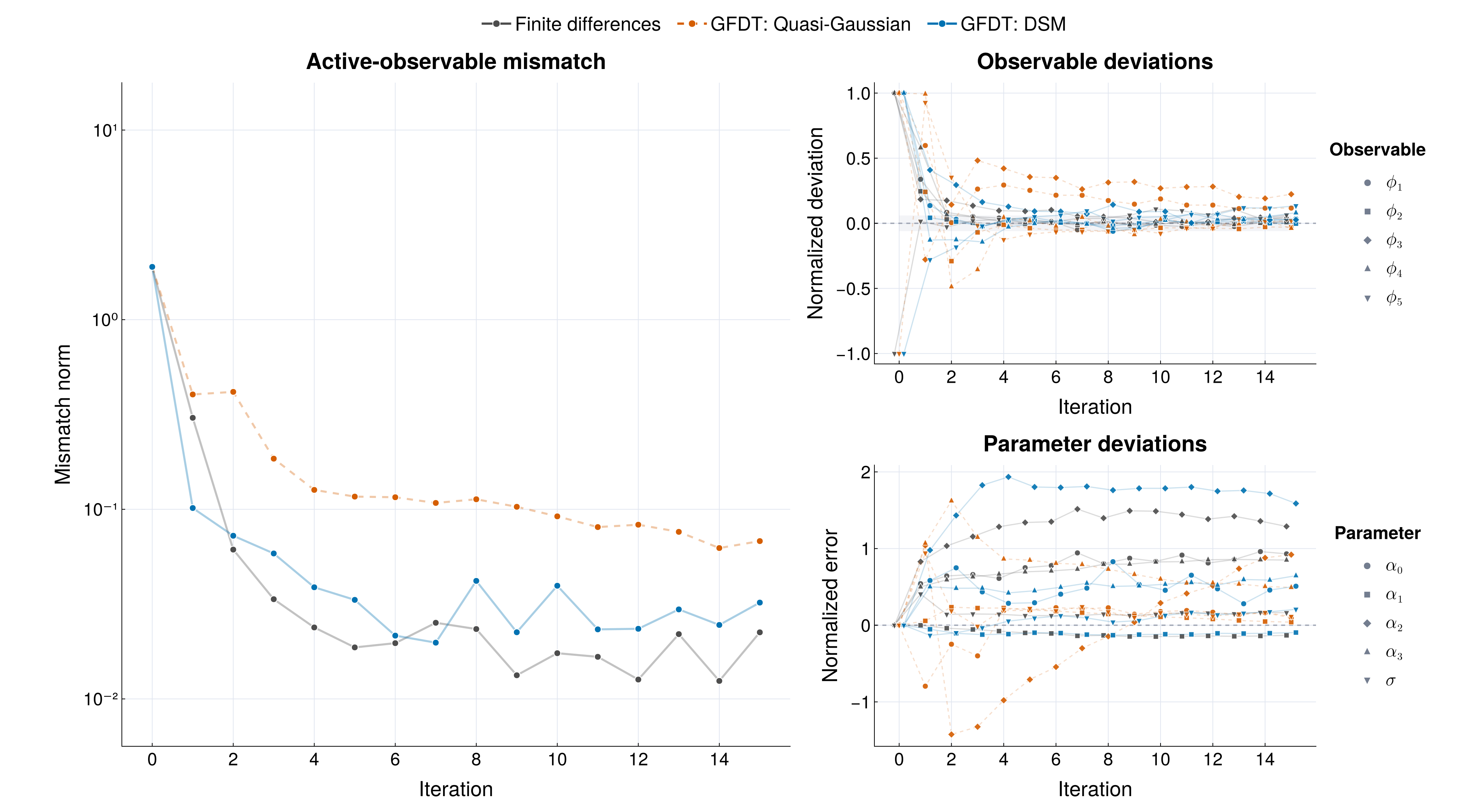}
  \caption[Lorenz--96 calibration.]{\textbf{Lorenz--96 Calibration.} Left panel: active-observable mismatch norm on a logarithmic scale versus iteration. Top-right panel: normalized observable deviations for the five matched statistics $(\phi_1,\phi_2,\phi_3,\phi_4,\phi_5)=(\mu,V,\operatorname{Sk},\operatorname{Ku},C_1)$, each rescaled by its initial deviation from the target. Bottom-right panel: normalized parameter deviations for the five closure parameters $(\alpha_0,\alpha_1,\alpha_2,\alpha_3,\sigma)$, expressed as relative changes from the initial values. Colored curves: Finite differences (grey), GFDT: Quasi-Gaussian (orange), and GFDT: DSM (blue). The DSM-based GFDT and finite-difference branches exhibit comparable mismatch reduction and observable convergence, whereas the quasi-Gaussian branch decreases more slowly and retains larger residual bias, particularly in $\phi_1$ (mean) and $\phi_3$ (skewness).}
  \label{fig:l96_calibration_convergence_publication}
\end{figure*}

\medskip
\noindent\textbf{Results: Calibration convergence.}
Figure~\ref{fig:l96_calibration_convergence_publication} summarizes the outcome of the iterative calibration. The left panel tracks the active-observable mismatch norm on a logarithmic scale over 15 iterations. The top-right panel shows the normalized observable deviations for the five matched statistics $(\phi_1,\phi_2,\phi_3,\phi_4,\phi_5)=(\mu,V,\operatorname{Sk},\operatorname{Ku},C_1)$, each rescaled by its initial deviation from the target, and the bottom-right panel shows the normalized parameter deviations for the five closure parameters $(\alpha_0,\alpha_1,\alpha_2,\alpha_3,\sigma)$, expressed as relative changes from the initial values. The DSM-based GFDT (using the periodic U-Net score representation) and finite-difference branches follow qualitatively similar trajectories and exhibit comparable mismatch reduction, whereas the quasi-Gaussian branch decreases more slowly and retains larger residual biases, most visibly in $\phi_1$ (mean) and $\phi_3$ (skewness).

The left panel of Figure~\ref{fig:l96_calibration_convergence_publication} condenses the multivariate convergence into the active-observable mismatch norm, a single scalar metric that measures the overall discrepancy between the model statistics and their targets. The most pronounced separation occurs after the first iteration: the DSM-based GFDT branch exhibits the steepest initial reduction, after which it remains in the same low-error regime as the finite-difference reference. The quasi-Gaussian branch decreases more slowly and stays at a substantially larger residual throughout the 15 iterations.

\medskip
\noindent\textbf{Computational perspective.}
These results illustrate the practical trade-off between the three Jacobian strategies. The non-Gaussian GFDT calibration substantially outperforms the quasi-Gaussian surrogate and approaches the finite-difference reference, while retaining a more favorable computational structure. Once a score representation has been specified, both GFDT variants obtain the entire $5\times5$ Jacobian from one unperturbed trajectory, whereas finite differences requires two perturbed integrations for each of the five parameters (ten perturbed runs per iteration for central differences), so its cost grows linearly with the parameter dimension. Finite differences also requires a careful choice of perturbation step size: if the perturbation is too small, the differenced observable means are swamped by sampling noise unless much longer trajectories are used, whereas overly large perturbations leave the linear-response regime and bias the estimated Jacobian. The GFDT formulation avoids this step-size dilemma entirely, since it computes an analytic (infinitesimal) derivative rather than a discrete difference.

\section{\label{sec:conclusions} Conclusions}

This paper has introduced a response-theoretic framework that recasts parameter calibration of ergodic stochastic models as a fluctuation--dissipation problem. Our central result is that each entry of the sensitivity matrix $\bm{S}=\partial\langle\bm{\mathcal{A}}\rangle/\partial\bm{\theta}$ admits an exact linear-response representation as a time-correlation integral evaluated along a single unperturbed trajectory, involving the observable, the parameter derivatives of the drift and diffusion, and the score function $\bm{s}=\nabla_{\bm{x}}\log\rho$ of the invariant density. This yields both a conceptual and a computational alternative to finite-difference, adjoint, and ensemble-based sensitivity analysis: the full Jacobian can be obtained without perturbed simulations, intrusive code modifications, or repeated model integrations. When combined with data-driven score estimation (KGMM in low dimensions and a periodic convolutional U-Net in higher dimensions), the framework remains accurate for non-Gaussian steady states and extends naturally to discontinuous observables, including exceedance indicators.

We validated the approach across a hierarchy of models of increasing complexity. For the Ornstein--Uhlenbeck process, the GFDT recovers the exact analytic sensitivity of the stationary variance with respect to the damping parameter. For the quartic double-well potential, the GFDT with the exact score closely matches high-accuracy finite-difference reference Jacobians, including the sensitivity of a tail exceedance probability, while a quasi-Gaussian closure produces incorrect signs and magnitudes for the most non-Gaussian observables. For the scalar stochastic climate model and the slow--fast ENSO triad with multiplicative noise, embedding GFDT Jacobians (computed via KGMM) in a regularized Gauss--Newton loop recovers the true parameters from biased initial guesses, with convergence trajectories that closely track the finite-difference baseline. In the most challenging test, stochastic closure calibration for the two-scale Lorenz--96 system ($K=36$ resolved variables, five closure parameters, and five matched observables), the DSM-based GFDT branch achieves the strongest initial decrease of the observable mismatch and converges to a low-error regime comparable to the finite-difference reference, whereas the quasi-Gaussian branch remains trapped at a significantly larger residual.

From a practical standpoint, the approach integrates naturally into existing simulation workflows and is especially attractive when forward simulation is the dominant computational bottleneck. Once a baseline trajectory is available, the score is estimated from that trajectory and can then be reused across different calibration targets and observable sets, amortizing the training cost. Each Gauss--Newton iteration requires only one forward integration (plus the score fitting and correlation evaluation), regardless of the number of parameters, whereas finite differences require two perturbed integrations per parameter. The Lorenz--96 results demonstrate that this computational structure extends beyond low-dimensional examples: with a symmetry-aware score architecture, the same response-theoretic machinery can calibrate stochastic parameterizations for multiscale chaotic systems.

Several limitations should be acknowledged. First, the GFDT provides a first-order (linear) approximation to the parameter-to-statistics map; when the true map is strongly nonlinear, the resulting tangent approximation may have a limited radius of validity, requiring smaller step sizes or trust-region safeguards in the optimization loop. Second, the method assumes that the unperturbed dynamics are sufficiently mixing for the two-time correlations to decay and for the time integrals in the GFDT to converge; weakly ergodic systems with long-lived metastable states or slow spectral gaps would require proportionally longer trajectories. Third, the accuracy of the sensitivity matrix depends directly on the quality of the estimated score and, for diffusion-parameter sensitivities, on its spatial derivatives (the score Jacobian or divergence); errors in these estimates propagate into the correlation integrals and can degrade the calibration, as seen in the weaker agreement in the $\partial/\partial\sigma$ column of the Lorenz--96 Jacobian. Fourth, identifiability constraints may limit which parameters can be tuned simultaneously: in the scalar climate model, for example, attempting to calibrate all five parameters against only three observables led to near-degeneracy, requiring a subset to be held fixed. Importantly, most of these limitations reflect the general structure of local calibration problems in complex dynamical systems rather than restrictions peculiar to the present method, and the corresponding regularity, identifiability, and mixing assumptions are met in many practically relevant settings.

These considerations point to several directions for future work. Higher-order response formulas or trust-region extensions could enlarge the domain of validity beyond the linear-response regime, enabling larger parameter steps per iteration. Non-stationary and cyclo-stationary formulations of the GFDT (the general derivation in App.~\ref{sec:derivation_gfdt} already accommodates time-dependent baselines) would open the door to calibrating parameters of systems with seasonal forcing or transient dynamics. On the score estimation side, structure-preserving and physics-informed architectures could improve scalability to higher-dimensional state spaces, while variance-reduced estimators for the correlation integrals (e.g., control-variate or multifidelity strategies) could sharpen the sensitivity estimates from shorter trajectories. A particularly promising direction is the calibration of stochastic models whose drift and diffusion are themselves parameterized by neural networks. In such neural-SDE settings, the number of trainable parameters can be very large, making repeated perturbed integrations especially costly and rendering favorable scaling with parameter dimension particularly valuable when the calibration target is defined through stationary or long-time statistics rather than pathwise losses. Finally, the GFDT-derived sensitivity matrix carries information about the local geometry of the parameter-to-statistics map that could be exploited for uncertainty quantification: response-based Fisher information metrics, for instance, could provide approximate posterior covariances within Bayesian calibration pipelines, connecting the present framework to principled probabilistic inference. More broadly, the connection between response theory and parameter calibration established here suggests that the rich mathematical structure of fluctuation--dissipation relations---developed over decades in nonequilibrium statistical mechanics---may have a substantially wider role to play in scientific inverse problems and data-driven model development than has previously been recognized.

\begin{acknowledgments}
We acknowledge support from Schmidt Sciences, BC3, the Aeolus Labs Research Residency Program, and the Altamira Collaboratory.
\end{acknowledgments}

\appendix

\section{Derivation of the GFDT}
\label{sec:derivation_gfdt}

Consider an SDE with baseline drift/diffusion and small time-dependent perturbations
\begin{align}
d\bm{x}_t
= \left[\bm{F}(\bm{x}_t,t) + \varepsilon\,\bm{\Psi}(\bm{x}_t,t)\right]\,dt
+ \left[\bm{\Sigma}(\bm{x}_t,t) +\varepsilon\,\bm{\Lambda}(\bm{x}_t,t)\right]\,d\bm{W}_t.
\end{align}
The Fokker–Planck equation for the density $\rho(\bm{x},t)$ is
\begin{align}
  \partial_t \rho
  + \nabla \cdot \left(\left[\bm{F}+ \bm{\Psi}\right] \rho - \frac{1}{2} \nabla \cdot \left[\left(\bm{\Sigma}+\varepsilon\,\bm{\Lambda}\right)\left(\bm{\Sigma}+\varepsilon\,\bm{\Lambda}\right)^T \rho\right]\right) = 0.
\end{align}
We write the density as $\rho = \rho_0 + \varepsilon \rho_1$ where we chose $\rho_0$ to satisfy, 
\begin{align}
  \label{eq:fp_unperturbed}
  \partial_t \rho_0 + \nabla \cdot \left(\bm{F}\,\rho_0 - \frac{1}{2} \nabla \cdot \left[\bm{\Sigma}\bm{\Sigma}^T\,\rho_0\right]\right) = 0.
  \end{align}
The equation for $\rho_1$ is then (after dividing by $\varepsilon$), 
  \begin{align}
    \label{eq:fp_perturbed}
  \partial_t \rho_1
  + \nabla \cdot \left(\bm{F}\rho_1 - \frac{1}{2} \nabla \cdot \left[\bm{\Sigma} \bm{\Sigma}^T \rho_1\right]\right) = -\nabla \cdot \left(\tilde{\bm{F}} \rho_0 \right)  -\varepsilon \nabla \cdot \left(\tilde{\bm{D}} \rho_1 \right)
  \end{align}
where
\begin{align}
\tilde{\bm{F}} = \bm{\Psi}
- \frac{1}{2}\nabla \cdot \left( \bm{\Sigma}\bm{\Lambda}^T + \bm{\Lambda}\bm{\Sigma}^T  + \varepsilon \bm{\Lambda}\bm{\Lambda}^T \right)
- \frac{1}{2}\left( \bm{\Sigma}\bm{\Lambda}^T + \bm{\Lambda}\bm{\Sigma}^T + \varepsilon \bm{\Lambda}\bm{\Lambda}^T \right)\nabla \ln \rho_0
\\
\tilde{\bm{D}} = \bm{\Psi}
- \frac{1}{2}\nabla \cdot \left( \bm{\Sigma}\bm{\Lambda}^T + \bm{\Lambda}\bm{\Sigma}^T  + \varepsilon \bm{\Lambda}\bm{\Lambda}^T \right)
- \frac{1}{2}\left( \bm{\Sigma}\bm{\Lambda}^T + \bm{\Lambda}\bm{\Sigma}^T + \varepsilon \bm{\Lambda}\bm{\Lambda}^T \right)\nabla \ln \rho_1
\end{align}
Equation~\ref{eq:fp_perturbed} is exact; thus far no approximations have been made. We now make approximations in order to obtain the GFDT. 

Neglecting the $\mathcal{O}(\varepsilon)$ term in Equation~\ref{eq:fp_perturbed} inspires the following equation for the first-order correction to the density:
\begin{align}
\label{eq:fp_perturbed_approx}
\partial_t q
+ \nabla \cdot \left(\bm{F}q - \frac{1}{2} \nabla \cdot \left[\bm{\Sigma} \bm{\Sigma}^T q \right]\right) = \nabla \cdot \left(\bm{\tilde{\Psi}}  \rho_0 \right)
\end{align}
where 
\begin{align}
  \bm{\tilde{\Psi}}  = \bm{\Psi}
- \frac{1}{2}\nabla \cdot \left( \bm{\Sigma}\bm{\Lambda}^T + \bm{\Lambda}\bm{\Sigma}^T  \right)
- \frac{1}{2}\left( \bm{\Sigma}\bm{\Lambda}^T + \bm{\Lambda}\bm{\Sigma}^T \right)\nabla \ln \rho_0 .
\end{align}
The hope is that the solution to Equation~\ref{eq:fp_perturbed_approx} is a good approximation to the solution to Equation~\ref{eq:fp_perturbed}, i.e., $q = \rho_1 + \mathcal{O}(\varepsilon)$ as $\varepsilon \rightarrow 0$. Crucially, the operator on the left-hand side of Equation~\ref{eq:fp_perturbed_approx} is the same as the operator on the left-hand side of Equation~\ref{eq:fp_unperturbed}, i.e., the unperturbed Fokker--Planck operator.

Letting $G_0(\bm{x},\bm{y},t,s)$ be the Green’s function for the unperturbed Fokker–Planck operator,  we write the solution to Equation \ref{eq:fp_perturbed} as
\begin{align}
q(\bm{x},t)
&= -\, \int ds\, d\bm{y}\; G_0(\bm{x},\bm{y},t,s)\, \mathcal{B}(\bm{y},s)\, \rho_0(\bm{y},s),
\end{align}
where 
\begin{align}
\mathcal{B}(\bm{x},t) &\equiv \frac{\nabla \cdot \left(\bm{\tilde{\Psi}}(\bm{x},t)\,\rho_0(\bm{x},t)\right)}{\rho_0(\bm{x},t)} = \nabla \cdot \bm{\tilde{\Psi}} + \bm{\tilde{\Psi}} \cdot \nabla \ln \rho_0
\end{align}
For any observable $\mathcal{A}(\bm{x})$, the change in expectation of the observable due to the perturbation is approximated by
\begin{align}
\delta \langle \mathcal{A} \rangle &=  \varepsilon \int d\bm{x}\,\mathcal{A}(\bm{x})\,q(\bm{x},t) \\
&= -\,\varepsilon \int d\bm{x}\,ds\,d\bm{y}\; \mathcal{A}(\bm{x})\, G_0(\bm{x},\bm{y},t,s)\, \mathcal{B}(\bm{y},s)\, \rho_0(\bm{y},s),
\end{align}
which we expect to be first order accurate in $\varepsilon$. 
The integral
\begin{align}
\mathcal{K}(t, s ) = -\int d\bm{x}\,d\bm{y}\; \mathcal{A}(\bm{x})\, G_0(\bm{x},\bm{y},t,s)\, \mathcal{B}(\bm{y},s)\, \rho_0(\bm{y},s),
\end{align}
represents a temporal autocorrelation of observable $\mathcal{A}$ at time $t$ with the observable $\mathcal{B}$ at time $s$, i.e. 
\begin{align}
\mathcal{K}(t, s) = -\langle \mathcal{A}(x_t) \mathcal{B}(x_s, s) \rangle
\end{align}
with respect to the unperturbed dynamics. We approximate the temporal autocorrelation with the empirical estimate
\begin{align}
\mathcal{K}(t,s) = -\frac{1}{N}\sum_{\omega=1}^{N} \mathcal{A}(\bm{x}_{\omega}(t))\, \mathcal{B}(\bm{x}_{\omega}(s),s).
\end{align}
where $\omega$ indexes the ensemble of $N$ trajectories $\{\bm{x}_\omega(t)\}$. This observation comes from the fact that $G_0(\bm{x},\bm{y},t,s) \rho_0(\bm{y}, s)$ represents the joint probability density of the unperturbed trajectory, $\rho_J$, of$(\bm{x}_s, \bm{x}_t)$ at times $(t, s)$, that is, 
\begin{align}
\rho_J(\bm{x}_t, \bm{x}_s, t, s) =  G_0(\bm{x}_t,\bm{x}_s,t,s) \rho_0(\bm{x}_s, s).
\end{align}
Ultimately, this leads  to the change in observable formula, 
\begin{align}
\delta \langle \mathcal{A} \rangle = \varepsilon \int_0^t ds\, \mathcal{K}(t,s)
\end{align}
In the derivation, observe that we did not make any assumptions about the baseline probability density $\rho_0$ or the form of the observable $\mathcal{A}$. In particular, $\rho_0$ can be a non-stationary distribution and $\mathcal{A}$ can be a discontinuous function of state space. 

\section{Technical Details for the Calibration Experiments}
\label{app:calib_details}

This appendix collects numerical details deferred from Sec.~\ref{sec:o_results}. In the scalar and triad experiments, the non-Gaussian score is estimated with KGMM \cite{giorgini2025kgmm}. Given equilibrium samples $\{\bm{\mu}_i\}_{i=1}^{N}$, we generate perturbed points
\begin{equation}
\bm{x}_i=\bm{\mu}_i+\sigma_G\bm{z}_i,
\qquad
\bm{z}_i\sim\mathcal{N}(\bm{0},\bm{I}),
\label{eq:kgmm_perturb_samples}
\end{equation}
cluster $\{\bm{x}_i\}$ into sets $\{\Omega_j\}_{j=1}^{N_C}$, and estimate the score at the cluster centroids $\bm{C}_j$ through
\begin{equation}
\widehat{\bm{s}}(\bm{C}_j)
=
-\frac{1}{\sigma_G}\,
\frac{1}{|\Omega_j|}\sum_{\bm{x}_i\in\Omega_j}\bm{z}_i .
\label{eq:kgmm_score_centroid}
\end{equation}
A smooth MLP interpolant of $\{(\bm{C}_j,\widehat{\bm{s}}(\bm{C}_j))\}$ then provides a differentiable score field on state space. In both low-dimensional experiments we use $\sigma_G=0.1$, choose $N_C$ according to the heuristic scaling $N_C\propto \sigma_G^{-d_{\mathrm{eff}}}$, and obtain the score Jacobian entering the diffusion-response formulas by reverse-mode automatic differentiation of the interpolant.

\subsection{Scalar Stochastic Model for Low-Frequency Variability}
\label{app:scalar_details}

The scalar calibration experiment of Subsec.~\ref{sec:reduced1d} updates the identifiable subset $(F,a,\sigma)$ while holding $(b,c)$ fixed. At each iteration, a single stationary trajectory of length $T=10^7$ with time step $\Delta t=0.01$ is used both to estimate the observable vector $(\langle x\rangle,\langle x^2\rangle,\langle \mathbf{1}_{x\le\beta}\rangle)$ and to assemble the GFDT correlation integrals. The active weight matrix in \eqref{eq:GN-step} is
\[
\bm{B}^{(k)}
=
\left(\widehat{\mathrm{Cov}}\!\big[\bm{\mathcal{A}}(x)\big]+10^{-10}I\right)^{-1},
\]
where the empirical covariance is computed from the same trajectory. Iterations are terminated when the mismatch norm $\|\bm{r}^{(k)}\|$ falls below $10^{-3}$ or when the number of iterations exceeds $10$.

For the non-Gaussian branch, the KGMM score estimator is fit to stationary samples extracted from the calibration trajectory. The centroid scores from \eqref{eq:kgmm_score_centroid} are interpolated with a fully connected MLP with hidden widths $(50,25)$, Swish activations, batch size $32$, and $2000$ Adam epochs. Because the state is one-dimensional, the score derivative needed for the $\sigma$-response is obtained exactly by automatic differentiation of this interpolant. The analytical score available for this model provides an additional benchmark, while the finite-difference reference is obtained from separate perturbed simulations and the Gaussian closure is built from the empirical mean and variance of the same baseline trajectory.

\subsection{Slow--Fast Triad Model and Application to ENSO}
\label{app:triad_details}

The triad experiment of Subsec.~\ref{sec:triad_enso} calibrates all six parameters $(d_u,\omega,d_{\tau},\sigma_1,\sigma_2,\sigma_3)$ against the six second-order observables $(u_1^2,u_2^2,u_1u_2,\tau^2,u_1\tau,u_2\tau)$. As in the scalar example, each Gauss--Newton step is based on a single stationary trajectory of length $T=10^7$ with time step $\Delta t=0.01$, which is reused for both observable estimation and GFDT response evaluation. The weight matrix again takes the covariance-regularized form
\[
\bm{B}^{(k)}
=
\left(\widehat{\mathrm{Cov}}\!\big[\bm{\mathcal{A}}(u_1,u_2,\tau)\big]+10^{-10}I\right)^{-1},
\]
and the stopping criterion is the same: iterations are terminated when the mismatch norm $\|\bm{r}^{(k)}\|$ falls below $10^{-3}$ or when the number of iterations exceeds $10$.

For the KGMM branch, the perturbed point cloud in \eqref{eq:kgmm_perturb_samples} is clustered in the three-dimensional state space $(u_1,u_2,\tau)$, and the centroid scores are interpolated with a fully connected MLP with hidden widths $(100,50)$, Swish activations, batch size $32$, and $200$ Adam epochs. Reverse-mode automatic differentiation yields the full $3\times 3$ score Jacobian needed for the multiplicative-noise response, so the diffusion sensitivities in \eqref{eq:triad_diff_conjugates} are evaluated without additional finite-difference approximations. The finite-difference baseline is obtained from perturbed integrations of the triad model, while the Gaussian closure uses the empirical Gaussian fitted to the same baseline trajectory.

\subsection{Stochastic Closure Calibration in the Lorenz--96 System}
\label{app:l96_details}

\medskip
\noindent\textbf{Target statistics and initialization.}
The target vector $\bm{A}$ and the initial parameter vector $\bm{\theta}^{(0)}$ are both derived from a single long equilibrium integration of the stochastic two-scale system \eqref{eq:l96_X}--\eqref{eq:l96_Y} over the interval $[0,5\times 10^5]$ with time step $\Delta t=0.005$. The five target observables in \eqref{eq:l96_forward_map_impl} are estimated from the slow variables along this trajectory. The same trajectory is also used to diagnose the unresolved tendency acting on the slow variables. For each site $k$ and time $t_n$, we write
\begin{equation}
R_{k,n}
=
F - X_k(t_n) - X_{k-1}(t_n)\bigl(X_{k-2}(t_n)-X_{k+1}(t_n)\bigr)
- \frac{X_k(t_{n+1})-X_k(t_{n-1})}{2\Delta t}.
\label{eq:l96_initial_fit_resolved}
\end{equation}
The drift coefficients are then determined by least-squares projection onto the local cubic ansatz
\begin{equation}
R_{k,n}\approx \alpha_0 + \alpha_1 X_k(t_n) + \alpha_2 X_k(t_n)^2 + \alpha_3 X_k(t_n)^3.
\label{eq:l96_initial_fit_cubic}
\end{equation}
The additive-noise amplitude is estimated from the residual process of \eqref{eq:l96_initial_fit_cubic} by integrating its empirical autocovariance up to the first non-positive lag and averaging over lattice sites. This construction follows the stochastic-parameterization strategy of Arnold et al.~\cite{arnold2013stochastic}: the conditional mean of the unresolved forcing is represented by a local cubic function of the resolved state, and the remaining fast variability is modeled by an additive stochastic term.

\medskip
\noindent\textbf{GFDT Jacobians and score surrogates.}
For the reduced stochastic model \eqref{eq:l96_stochastic}, the drift perturbation associated with $\alpha_p$ is local,
\[
\bigl(\partial_{\alpha_p} b_{\bm{\theta}}(X)\bigr)_k = -X_k^p,
\qquad
p=0,1,2,3,
\]
and the diffusion perturbation associated with $\sigma$ is $\partial_\sigma \bm{\Sigma}=\bm{I}$. The corresponding conjugate observables are listed in \eqref{eq:l96_g_alpha0}--\eqref{eq:l96_g_sigma}. At each calibration iteration, one unperturbed trajectory of the reduced model is integrated over $[0,5\times10^5]$ with time step $\Delta t=0.005$, and the first $6$ time units are discarded. The post-transient portion of this trajectory is used for three purposes: to estimate the observable vector entering the update, to fit the U-Net and Gaussian score surrogates from snapshots saved every $200\Delta t=1$, and to evaluate the GFDT response integrals from a block of $10^6$ samples at cadence $2\Delta t=0.01$. The correlation functions are computed by one-sided FFT estimators; after mean removal and conversion to step responses, they are truncated at large lags and averaged over the asymptotic window
\[
t\in[t_{\mathrm{start}},t_{\mathrm{end}}]=[2,5]
\]
to obtain the stationary Jacobian.

Three Jacobian constructions are compared. In the U-Net-based GFDT calculation, the score is learned from the post-transient trajectory, subsampled every $200\Delta t=1$, using a periodic one-dimensional convolutional U-Net with one input channel and one output channel, base width $32$, channel multipliers $(1,2)$, kernel size $5$, and group normalization with $8$ groups. At each calibration iteration, the network is warm-started from the previous iterate and trained for $20$ epochs with batch size $1024$, learning rate $8\times 10^{-4}$, and denoising level $0.025$. The divergence term in the diffusion conjugate observable of \eqref{eq:l96_g_sigma} is approximated by a Hutchinson trace estimator with $10$ Rademacher probes and finite-difference radius $0.03$. The quasi-Gaussian GFDT calculation replaces the score by that of the empirical Gaussian fitted to the same trajectory. The finite-difference reference Jacobian is computed by the central-difference approximation
\begin{equation}
S_{mj}^{\mathrm{FD},(n)}
\approx
\frac{
\widehat{\mathcal{G}}_{m}\!\left(\bm{\theta}^{(n)}+h_j\bm{e}_j\right)
-
\widehat{\mathcal{G}}_{m}\!\left(\bm{\theta}^{(n)}-h_j\bm{e}_j\right)
}{2h_j},
\label{eq:l96_fd_jacobian}
\end{equation}
where $\bm{e}_j$ is the $j$th Euclidean basis vector and $\widehat{\mathcal{G}}_{m}$ denotes the empirical stationary average of the observable $\Phi_m$ computed from a perturbed reduced-model trajectory. For each parameter $j$, the two empirical averages in \eqref{eq:l96_fd_jacobian} are obtained from trajectories integrated over $[0,216000]$ with time step $\Delta t=0.005$, discarding the first $6$ time units; common random numbers are used for the two perturbations. The parameter-wise step sizes satisfy
\[
h_i=\max\!\bigl(h_i^{\mathrm{abs}},\, h_{\mathrm{rel}}\max(|\theta_i|,1)\bigr),
\]
with $h^{\mathrm{abs}}=(0.05,0.02,0.01,0.002,0.05)$ and $h_{\mathrm{rel}}=5\times 10^{-3}$. If a perturbed trajectory becomes unstable, the corresponding step size is reduced automatically.

For the U-Net score estimate, we additionally apply the discrete score correction proposed in \cite{Giorgini2024}. In exact arithmetic, the stationary score satisfies the identity $-\langle X_i s_j(X)\rangle=\delta_{ij}$, but finite-sample estimation and network approximation introduce a small defect. Writing the empirical moment matrix in the form
\[
-\widehat{\mathbb{E}}\!\left[X\,s(X)^\top\right]=I-\Xi,
\]
we replace the raw score by the linearly recalibrated field
\[
\varsigma(X)=s(X)(I-\Xi)^{-1}.
\]
This correction enforces the discrete identity on the calibration ensemble and improves the short-lag response estimates entering the GFDT Jacobian.

\medskip
\noindent\textbf{Weighted update and safeguards.}
Let $S^{(n)}\in\mathbb{R}^{5\times 5}$ denote the Jacobian at iteration $n$, and let
\[
r^{(n)}=\bm{\mathcal{G}}(\bm{\theta}^{(n)})-\bm{A}.
\]
The weight matrix is taken to be
\[
\bm{B}^{(n)}=\mathrm{diag}\!\Big(\widehat{\mathrm{Var}}\big[\Phi_1\big]^{-1},\dots,\widehat{\mathrm{Var}}\big[\Phi_5\big]^{-1}\Big),
\]
where the empirical variances are computed from the current unperturbed reduced-model trajectory. Before solving the linearized system, each column of the Jacobian is rescaled to have unit $\bm{B}^{(n)}$-weighted norm. Writing
\[
D^{(n)}=\mathrm{diag}\!\Big(\|S^{(n)}_{\cdot 1}\|_{\bm{B}^{(n)}}^{-1},\dots,\|S^{(n)}_{\cdot 5}\|_{\bm{B}^{(n)}}^{-1}\Big),
\]
we define the induced regularization matrix
\[
\Gamma^{(n)}=\gamma\big(D^{(n)}\big)^{-2},
\qquad
\gamma=10^{-2},
\]
and compute the parameter increment from
\begin{equation}
\Big[\big(S^{(n)}\big)^\top \bm{B}^{(n)} S^{(n)} + \Gamma^{(n)}\Big]\delta^{(n)}
=
\big(S^{(n)}\big)^\top \bm{B}^{(n)} r^{(n)}.
\label{eq:l96_regularized_update}
\end{equation}
This form is equivalent to applying isotropic Tikhonov regularization in the equilibrated coordinates defined by $D^{(n)}$. The column equilibration balances the parameter-sensitivity scales after weighting by $\bm{B}^{(n)}$, improves the conditioning of the linearized system, and yields a regularization that is commensurate with the local scale of each parameter direction. Each calibration method advances with a single damped update
\[
\bm{\theta}^{(n+1)}=\bm{\theta}^{(n)}-\lambda^{(n)}\delta^{(n)}.
\]
Here $\lambda^{(n)}=1$ unless the column-equilibrated Jacobian becomes ill-conditioned, in which case the relative step size is capped by enforcing
\[
\frac{\|\lambda^{(n)}\delta^{(n)}\|_2}{\|\bm{\theta}^{(n)}\|_2}\le 5\times 10^{-2}
\qquad\text{whenever}\qquad
\operatorname{cond}\!\big(S^{(n)}D^{(n)}\big)\ge 10^4.
\]
The candidate parameter vector is subjected to short stochastic stability checks at both the score-fitting cadence and the GFDT cadence, and, if instability is detected, the update is retried with up to three successive step halvings. Every post-update evaluation of $\bm{\mathcal{G}}(\bm{\theta})$ uses observable averages computed from the same post-transient unperturbed trajectory. Since the Lorenz--96 closure problem is an imperfect-closure calibration problem, no stopping rule based on parameter convergence is imposed. Instead, the U-Net-based GFDT, quasi-Gaussian GFDT, and finite-difference iterations are all advanced for $15$ iterations, subject to the conditioning and stability safeguards described above. The resulting histories are those shown in Fig.~\ref{fig:l96_calibration_convergence_publication}.

\bibliography{apssamp}

\end{document}